\documentclass[10pt,preprint]{sigplanconf-eurosys}
\usepackage{epsfig,endnotes,algorithm,verbatim,algpseudocode}
\usepackage{amsmath}
\usepackage{balance}
\usepackage{flushend}
\usepackage{multirow}
\usepackage{color}
\usepackage{wrapfig}
\usepackage{graphicx}
\usepackage[font=small,labelfont=bf]{caption}
\usepackage{url}
\usepackage{subcaption}
\usepackage{booktabs}
\usepackage{mwe}
\usepackage{hyperref}
\usepackage{pifont}
\newcommand{\cmark}{\ding{51}}%
\newcommand{\xmark}{\ding{55}}%
\usepackage{times}
\usepackage{datetime}
\usepackage{url}

\usepackage[numbers]{natbib}

\usepackage{etoolbox}

 \newcommand{\squishlist}{
   \begin{list}{$\bullet$}
    { \setlength{\itemsep}{0pt}      \setlength{\parsep}{3pt}
      \setlength{\topsep}{3pt}       \setlength{\partopsep}{0pt}
      \setlength{\leftmargin}{1.5em} \setlength{\labelwidth}{1em}
      \setlength{\labelsep}{0.5em} } }

\newcommand{\squishend}{
    \end{list}  }

 \date{}

\authorinfo{Faria Kalim, Le Xu, Sharanya Bathey, Richa Meherwal, Indranil Gupta}{University of Illinois, Urbana Champaign\\\{kalim2, lexu1, bathey2, meherwa2, indy\}@illinois.edu}

\makeatletter
\def\@copyrightspace{\relax}
\makeatother

\begin{document}

\title{Henge: Intent-driven Multi-Tenant Stream Processing\thanks{This work was supported in part by the following grants: NSF CNS 1409416, NSF CNS 1319527, AFOSR/AFRL FA8750- 11-2-0084, and Yahoo! and a generous gift from Microsoft.}}
\maketitle

\begin{abstract}
We present Henge,  a system to support intent-based multi-tenancy in modern stream processing applications.  Henge supports multi-tenancy as a first-class citizen: everyone inside an organization can now submit their stream processing jobs to a single, shared, consolidated cluster. Additionally, Henge allows each tenant (job) to specify its own intents (i.e., requirements) as a Service Level Objective (SLO) that captures latency and/or throughput. In a multi-tenant cluster, the Henge  scheduler adapts continually to meet jobs' SLOs in spite of limited cluster resources, and  under dynamic input workloads.  SLOs are soft and are based on utility functions. Henge continually tracks SLO satisfaction, and when jobs miss their SLOs, it wisely navigates the state space to perform resource allocations in real time,  maximizing total system utility achieved by all jobs in the system. Henge is integrated in Apache Storm and we present experimental results using both production topologies and real datasets. 
\end{abstract}

\section{Introduction}
\label{introduction}


Modern stream processing systems  process continuously-arriving data streams in real time, ranging from Web data to social network streams. For instance, several companies use Apache Storm~\cite{storm_web} (e.g., Weather Channel, Alibaba, Baidu, WebMD, etc.),  Twitter uses Heron~\cite{heron}, LinkedIn relies on Samza~\cite{samza} and others use Apache Flink~\cite{flink}. These systems provide high-throughput and low-latency processing of streaming data from advertisement pipelines  (Yahoo! Inc. uses Storm for this), social network posts (LinkedIn, Twitter), geospatial data (Twitter), etc.

While stream processing systems for clusters have been around for decades~\cite{borealis, spade}, modern stream processing systems have scant support for {\it intent-based multi-tenancy}. We tease apart these two terms. First,  multi-tenancy allows multiple jobs to share a single consolidated cluster.  This capability is lacking in stream processing systems today--as a result,  many companies (e.g., Yahoo!) over-provision the stream processing cluster   and then physically apportion it among tenants (often based  on team priority). Besides higher cost, this entails manual administration  of multiple clusters and caps on allocation by the sysadmin, and manual monitoring of job behavior by each deployer. 

Multi-tenancy is attractive as it reduces acquisition costs and allows sysadmins to only manage a single consolidated cluster. Thus, this approach reduces capital and operational expenses (Capex \& Opex), lowers  total cost of ownership (TCO),  increases resource utilization, and allows jobs to elastically scale based on needs. Multi-tenancy has been explored for  areas such as key-value stores \cite{shue2012performance}, storage systems~\cite{wang2012cake}, batch processing \cite{yarn}, and others~\cite{mace2015retro}, yet it remains a vital need in modern stream processing systems. 

Second, we believe the deployer of each job should be able to clearly specify their performance expectations as an intent to the system, and it is the underlying engine's responsibility to meet this intent. This alleviates the developer's burden of monitoring and adjusting their job. Modern open-source stream processing systems like Storm~\cite{storm-multitenant-scheduler}  are very primitive and do not admit intents of any kind. 


Our approach is to allow each job in a multi-tenant environment to specify its intent as a Service Level Objective (SLO). The   metrics  in an SLO should be {\it user-facing}, i.e., understandable and settable by lay users such as a deployer who is not intimately familiar with the innards of the system. For instance, SLO metrics can capture latency and throughput expectations. SLOs do not include internal metrics like queue lengths or CPU utilization which can vary depending on the software, cluster, and job mix\footnote{However, these latter metrics can be monitored and used internally by the scheduler for self-adaptation.}. We believe lay users should not have to grapple with such complex metrics.

While there are myriad ways to specify SLOs (including potentially declarative languages paralleling SQL), our paper is best seen as one contributing {\it mechanisms} that are pivotal in order to build a truly intent-based distributed system for stream processing. In spite of their simplicity, our latency and throughput SLOs are immediately useful. Time-sensitive jobs (e.g., those related to an ongoing ad campaign) are latency-sensitive and can specify latency SLOs, while longer running jobs (e.g., sentiment analysis of trending topics) typically have throughput SLOs. 


\begin{table}[t]
\centering
\resizebox{0.50\textwidth}{!}{%
\begin{tabular}{l l l l l}
\toprule
Schedulers & Job Type & Adaptive & Reservation-Based & SLOs  \\ \hline
Mesos~\cite{mesos} & General & \xmark & \cmark (CPU, Mem, Disk, Ports) & \xmark \\ \hline
YARN~\cite{yarn} & General & \xmark & \cmark (CPU, Mem, Disk) & \xmark \\ \hline
Rayon~\cite{curino2014reservation} & Batch & \cmark  & \cmark (Resources across time) & \cmark \\ \hline
{\bf Henge} & {\bf Stream} & \cmark & \xmark \ {\bf (User-facing SLOs)} & \cmark \\ 
\bottomrule
\end{tabular}%
}
 \vspace{-0.3cm}
\caption{\bf Henge vs. Existing Schedulers.}
\vspace{-0.5cm}
\label{henge-comparison}

\end{table}


As Table~\ref{henge-comparison} shows, most existing schedulers use reservation-based approaches to specify intents: besides not being user-facing, these are very hard to estimate even for a job with a static workload~\cite{Morpheus}, let alone the dynamic workloads in streaming applications. 

We present Henge, a system  consisting of the first scheduler to support both multi-tenancy and per-job intents (SLOs) for modern stream processing engines. 
In a cluster of limited resources, Henge continually adapts  to meet jobs' SLOs in spite of other competing SLOs, both under natural system fluctuations, and under input rate changes due to diurnal patterns or sudden spikes. As our goal is to satisfy the SLOs of all jobs on the cluster, Henge must deal with the challenge of allocating resources to jobs continually and wisely. 



We implemented Henge into Apache Storm, one of the most popular modern open-source stream processing system. Our experimental evaluation uses real-world workloads: Yahoo! production Storm topologies, and Twitter datasets. We show that while satisfying SLOs, Henge prevents non-performing topologies from hogging cluster resources. It scales with cluster size and jobs, and is fault-tolerant.

This paper makes the following contributions: 
\squishlist
    \item We present the design of the Henge system and its state machine that manages resource allocation on the cluster.  
    \item We define a new throughput SLO metric called ``juice" and present an algorithm to calculate it.
    \item We define the structure of SLOs using utility functions. 
    \item We present implementation details of Henge's integration into Apache Storm.
    \item We present evaluation of Henge using real workloads.
\squishend
\section{Henge Summary}
\label{summary}

We now summarize key ideas behind our contributions. 

\noindent{\bf Juice:} As input rates can vary over time, it is infeasible for a throughput SLO to merely specify a desired absolute output rate value. Instead, we define a new {\it input rate-independent} metric for throughput SLOs called {\it juice}.  We show how Henge calculates juice for arbitrary topologies (Section~\ref{juice}). 

Juice lies in the interval [0, 1] and  captures the ratio of processing rate to input rate--a value of 1.0 is ideal and implies that the rate of incoming tuples equals rate of tuples being processed by the job. Throughput SLOs can then contain a minimum threshold for juice, making the SLO independent of input rate. We consider processing rate instead of output rate as this generalizes to cases where tuples may be filtered (thus they affect results but are never outputted themselves).

\noindent{\bf SLOs: } A job's SLO can capture either latency or juice (or a combination of both). The SLO contains: a) a threshold (min-juice or max-latency), and b) a {\it utility function}, inspired by soft real-time systems \cite{K11}.  The utility function maps current achieved performance (latency or juice) to a value which  represents the benefit to the job, even if it does not meet its SLO threshold. The function thus captures the developer intent that a job  attains full ``utility'' if its SLO threshold is met and partial benefit if not. We support monotonic utility functions: the closer the job is to its SLO threshold, the higher its achieved maximum possible utility. (Section~\ref{job-utilities}).%

 \noindent\textbf {State Space Exploration: }  At its core, Henge decides wisely how to change resource allocations of jobs (or rather of their basic units, operators) using a new {\it state machine} approach (Section~\ref{policy}). Our state machine is unique as it is {\it online} in nature: it takes one step at a time, evaluates its effect, and then moves on. This is a good match for unpredictable and dynamic scenarios such as modern stream processing clusters.
 
 
 
 The primary actions in our state machine are: 1) Reconfiguration (give resources to jobs missing SLO), 2) Reduction (take resources away from overprovisioned jobs satisfying  SLO), and 3) Reversion (give up an exploration path and revert to past high utility configuration). Henge  takes these actions  wisely. Jobs are given more resources as a function of the amount of congestion they face. Highly intrusive actions like reduction are minimized in number and frequency. 
 
 \noindent{\bf Maximizing System Utility: } Design decisions in Henge are aimed at  converging each job quickly to its maximum achievable utility in a minimal number of steps.
Henge attempts to maximize total achieved utility summed across all jobs. It does so by finding SLO-missing topologies, then their congested operators, and gives the operators thread resources according to their  congestion levels. Our approach creates a weak form of Pareto efficiency~\cite{wiki:pareto-efficiency}; in a system where jobs compete for resources, Henge transfers resources among jobs only if this will cause the system's utility to rise. 


\noindent\textbf {Preventing Resource Hogging:} Topologies with stringent SLOs may try to take over all the resources of the cluster.  To mitigate this, Henge  prefers giving resources to those topologies that: a) are farthest from their SLOs, and b) continue to show utility improvements due to recent Henge actions. This spreads resource allocation across all wanting jobs and prevents starvation and resource hogging.


\begin{table*}[ht!]
	\small
	\begin{tabular}{p{0.09\linewidth} p{0.4\linewidth} p{0.45\linewidth}}
		\toprule
		\textbf{Business} & \textbf{Use Case} & \textbf{SLO Type} \\
		\midrule
		\multirow{2}{6em}{The Weather Channel} & Monitoring natural disasters in real-time & Latency e.g., a tuple must be processed within 30 seconds \\ 
		& Processing collected data for forecasts 
		& Throughput e.g, processing data as fast as it can be read \\
		\midrule
		\multirow{2}{7em}{WebMD}  & Monitoring blogs to provide  real-time updates & Latency e.g., provide updates within 10 mins  \\ 
		& Search Indexing & Throughput e.g., index all new  sites at the rate they're found  \\
		\midrule
		\multirow{2}{7em}{E-Commerce Websites}  & Counting ad-clicks & Latency e.g., click count should be updated every second  \\ 
		&  Alipay uses Storm to process 6 TB logs per day & Throughput e.g., process logs at the rate of generation \\
		\bottomrule
	\end{tabular}
	\normalsize
	\vspace{-0.5em}
	\captionof{table}{\bf Stream Processing Use Cases and Possible SLO Types.}
	\label{table-slos}
	\vspace{-1.5em}
\end{table*}

The rest of the paper is organized as follows: Section~\ref{background} presents  background. Section~\ref{system-design} discusses core Henge design: SLOs and utilities (Section~\ref{job-utilities}), operator congestion (Section~\ref{congestion}), and the state machine (Section~\ref{policy}). Section~\ref{juice} describes juice and its calculation. Implementation details are in Section~\ref{system-architecture}, evaluation in Section~\ref{evaluation}, and related work in Section~\ref{related-work}. We conclude in Section~\ref{conclusion}.
\vspace{-0.2cm}
\section {Background}
\label{background}

\begin{figure}[!h]
\vspace{-0.2cm}
	\begin{center}
		\includegraphics[width=0.5\textwidth]{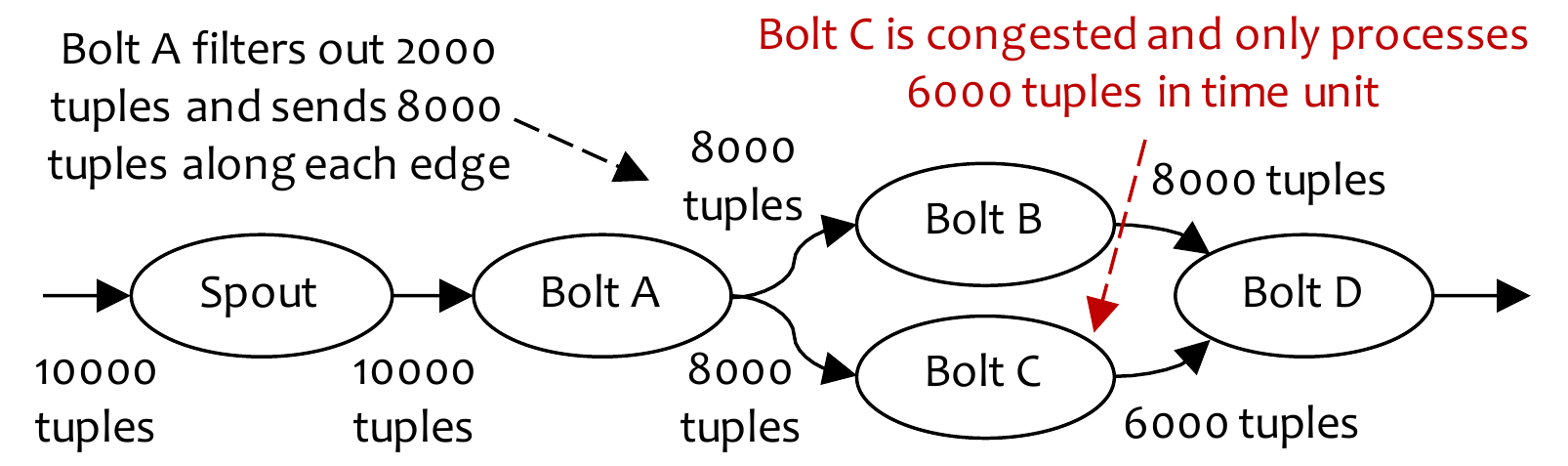}
	\end{center}
	\vspace{-1.5em}
	\caption{{\bf Sample Storm topology.} \it Showing tuples processed per unit time. Edge labels indicate number of tuples sent out by the parent operator to the child. (Congestion described in Section~\ref{juice}.) }
	\label{simple-topology}
	\vspace{-0.2cm}
\end{figure}

A stream processing job can be logically interpreted as a \emph{topology}, i.e., a directed acyclic graph of \emph{operators} (we sometimes use the Storm term ``bolt"). We use the terms job and topology interchangeably in this paper. An operator is a logical processing unit that applies user-defined functions on a stream of \emph{tuples}. Source operators (called spouts) pull input tuples while sink operators spew output tuples. The sum of output rates of sinks in a topology is its {\it output rate}, while the sum of all spout rates is the {\it input rate}. Each operator is  parallelized via multiple \emph{tasks}. Fig.~\ref{simple-topology} shows a topology with one spout and one sink.

 We consider long-running stream processing topologies with a continuous operator model. A topology is run on one or more worker processes, which in turn instantiate {\it executors} (threads), which run tasks specific to one operator.  An operator processes  streaming data one tuple at a time and forwards the tuples to the next operators in the topology.  Systems that follow such a model include Apache Storm~\cite{toshniwal2014storm}, Heron~\cite{heron}, Flink~\cite{flink} and Samza~\cite{samza}. 

\noindent\textbf{Definitions:} 
The latency of a tuple is the time between it entering  the topology from the source, to producing an output result on any sink.  A {\it topology's latency} is then the average of tuple latencies, measured over a period of time. A {\it topology's throughput} is  the number of tuples it processes per unit time. 

A {\it Service Level Objective (SLO)}~\cite{wiki:SLO} is a customer topology's requirement, in terms of latency and/or throughput. 
Table~\ref{table-slos} shows several real streaming applications~\cite{alipay-stats}, and the latency or throughput SLOs they may require.

Other examples of latency-sensitive jobs include applications that perform real-time analytics or real-time natural language processing, provide moderation services for chat rooms, count bid requests, or calculate real-time trade quantities  in stock markets. Examples of throughput-sensitive application include jobs that perform incremental checkpointing, count online visitors, or perform sentiment analysis.

\vspace{-0.5em}
\section{System Design}
\label{system-design}

We discuss Henge's utility functions (Section~\ref{job-utilities}), congestion metric (Section~\ref{congestion}), and its state machine (Section~\ref{policy}).

\vspace{-0.25cm}
\subsection{SLOs and Utility Functions}
\label{job-utilities}

Each topology's SLO contains: a) an SLO threshold (min-juice or max-latency), and b) a {utility function}. 
The utility function maps the current performance metrics of the job (i.e. its SLO metric) to a \emph{current} utility value. This approach abstracts away the type of SLO metric each topology has, and allows the scheduler to compare utilities across jobs. 


Currently, Henge supports both latency and throughput metrics in the SLO. Latency was defined in Section~\ref{background}. 
For throughput, we use a new SLO metric called juice which we  define concretely later in Section~\ref{juice} (for the current section, an abstract throughput metric suffices). 




 When the  SLO threshold cannot be satisfied, the job still desires {\it some} level of performance close to the threshold. Hence, utility functions must be monotonic--for a job with a latency SLO, the utility function must be monotonically non-increasing as latency rises, while for a job with a throughput SLO, it must be monotonically non-decreasing as throughput rises. 

Each utility function has a {\it maximum utility} value, achieved only when the SLO threshold is met e.g.,  a job with an SLO threshold of 100 ms would achieve its maximum utility  only if its current latency is below 100 ms. As latency grows above 100 ms, utility can fall or plateau but can never rise.

\begin{figure}[ht!]
	\begin{center}
		\includegraphics[width=1.0\columnwidth]{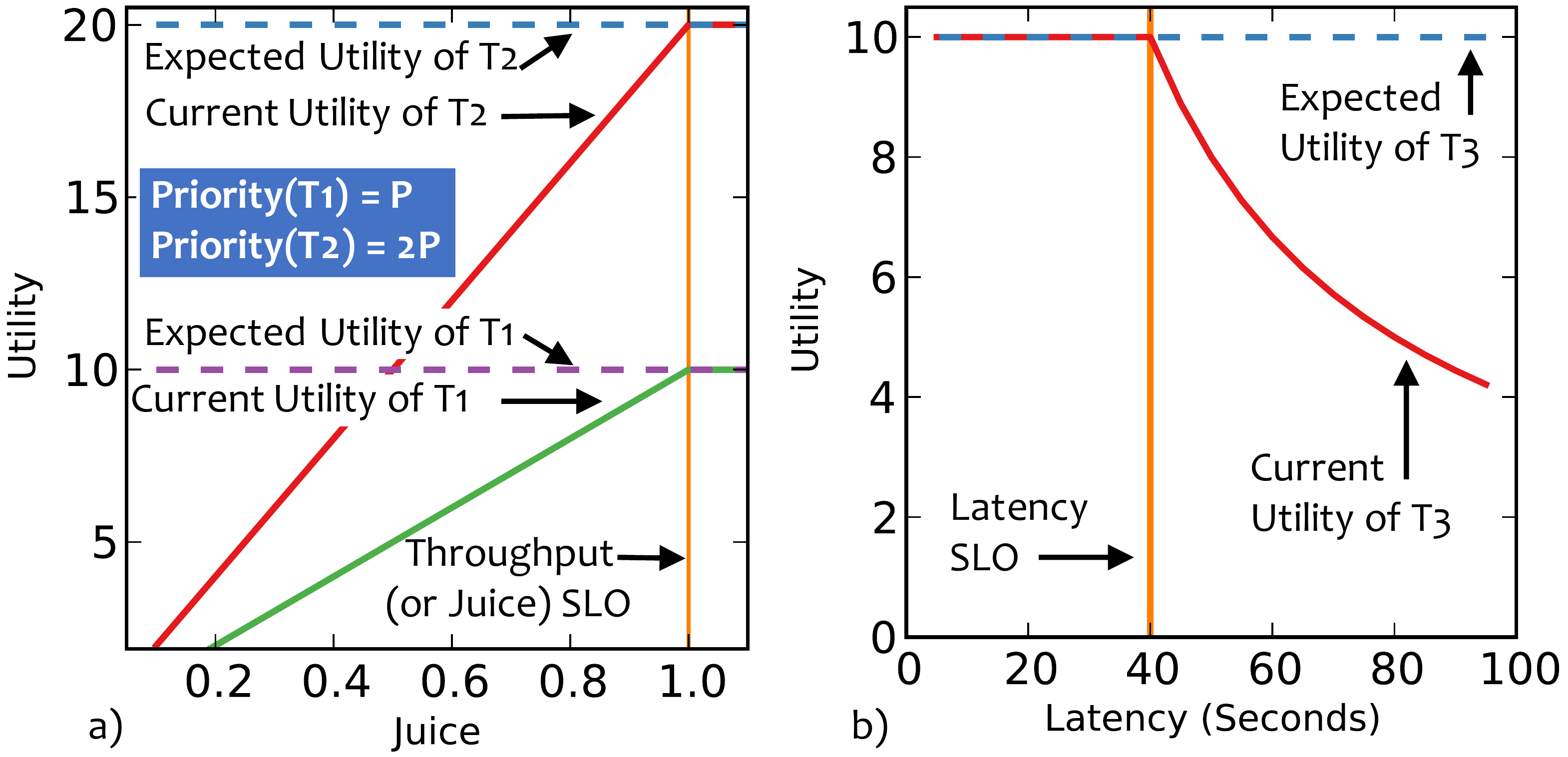}
	\end{center}
	\vspace{-0.4cm}
	\caption{ {\bf Knee Utility functions}. \it  (a) Throughput SLO utility, (b) Latency SLO utility.  } 
	\vspace{-0.4cm}
	\label{utility}
\end{figure}

The {maximum  utility value} is based on job priority.  For example, in Fig.~\ref{utility}a, topology T2 has twice the priority of T1, and thus has twice the maximum utility (20 vs. 10).

Given these requirements, Henge is able to allow a wide variety of shapes for its utility functions including: linear, piece-wise linear, step function (allowed because utilities are monotonically non-increasing instead of monotonically decreasing), lognormal, etc. Utility functions do not need to be continuous. All in all, this offers users flexibility in shaping utility functions according to individual needs. 




The concrete utility functions used in our Henge implementation are {\it knee} functions, depicted in Fig.~\ref{utility}. A knee function has two  pieces: a plateau beyond the SLO threshold, and a sub-SLO part for when the job does not meet the threshold. Concretely, the  achieved utility for jobs with throughput and latency SLOs respectively, are:

\vspace{-2em}

\small
\begin{center}
	\begin{equation} 
	\frac{\tt Current\ Utility}{Job\ Max\ Utility} = min(1, \frac{Current\ Throughput\ Metric}{SLO\ Throughput\   Threshold})
	\label{eq:5}
	\end{equation}
	\vspace{-0.5em}
	\begin{equation} 
	\label{eq:6}
	\frac{\tt Current\ Utility}{Job\ Max\ Utility} = min(1, \frac{SLO\ Latency\ Threshold}{Current\ Latency}) 
	\end{equation}
\end{center}
\normalsize

The sub-SLO is the last term inside ``min". 

For throughput SLOs, the sub-SLO is linear and arises from the  origin point. For latency SLOs, the sub-SLO is hyperbolic ($y \propto \frac{1}{x}$), allowing increasingly smaller utilities as latencies rise. Fig.~\ref{utility} shows a throughput SLO  (Fig.~\ref{utility}a) vs. latency SLO  (Fig.~\ref{utility}b). 


We envision Henge to be used internally inside companies, hence job priorities are set in a consensual way (e.g., by upper management). The utility function approach is also amenable to use in contracts like Service Level Agreements (SLAs), however these are beyond the scope of this paper.


\subsection{Operator Congestion Metric}
\label{congestion}

A topology misses its SLOs when some of its operators become \emph{congested}, i.e., have insufficient resources. To detect congestion our implementation uses a metric called operator \emph{capacity}~\cite{capacity-metric}.  However, Henge can also use  other existing congestion metrics, e.g.,  input queue sizes or ETP~\cite{xu:16:ic2e}.

Operator capacity captures the fraction of time that an operator spends processing tuples during a time unit. Its values lie in the range $[0.0, 1.0]$. If an executor's capacity is near 1.0, then it is close to being congested. 

Consider an  executor $E$ that runs several (parallel) tasks of a topology operator. Its capacity   is calculated as:
	 

\vspace{-0.6cm}
\begin{center}
\small

	\begin{equation} \label{eq:3}
	Capacity_{E} =  \frac{{Executed\ Tuples}_{E}\times{Execute\ Latency}_{E}}{Unit\ Time}
	\end{equation}
\end{center}
\vspace{-0.1cm}
\normalsize

\noindent where $Unit\ Time$ is a time window. The numerator multiplies the number of tuples executed in this window and their average execution latency to calculate the total time spent in executing those tuples.  
The operator capacity is then the maximum capacity across all executors containing it.

Henge considers an operator to be congested if its capacity is above the threshold of 0.3. This increases the pool of possibilities, as more operators  become candidates for receiving resources (described next). 

\subsection{Henge State Machine}
\label{policy}

\begin{figure}[h]
	\vspace{-0.1cm}
	\begin{center}
		\includegraphics[width=1\columnwidth]{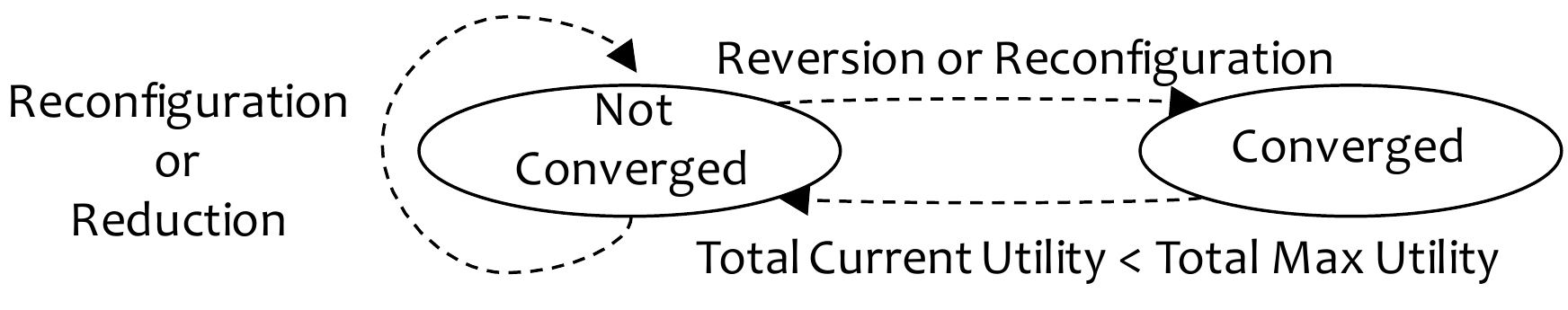}
	\end{center}
	\vspace{-0.3cm}
	\caption{\bf Henge's State Machine for the Cluster.}
	\label{henge-state-machine}
	\vspace{-0.4cm}
\end{figure}

The state machine (shown in Fig.~\ref{henge-state-machine}) considers all jobs in the cluster as a whole and wisely decides how many resources to give to congested jobs in the cluster and when to stop. The state machine is for the entire cluster, not per job. 

 The cluster is in  the Converged state if and only if either: a) all topologies have reached their maximum utility (i.e., satisfy their respective SLO thresholds), or b) Henge recognizes that no further actions will improve the performance of any topology, and thus it has reverted to the last best configuration. All other states are Not Converged. 
 
 To move among these two states, Henge uses three actions: Reconfiguration, Reduction, and Reversion.
 
 \subsubsection{Reconfiguration}

In the Not Converged state, a Reconfiguration gives resources to topologies missing their SLO. Reconfigurations occur in {\it rounds} which are periodic intervals (currently 10 s apart).  
In each round, Henge first sorts all topologies missing their SLOs, in descending order of their maximum utility, with ties broken by preferring lower current utility. It then picks the head of this sorted queue to allocate resources to. 
 This greedy strategy works best to maximize cluster utility. 

 Within this selected topology, the intuition is to increase each congested operator's resources by an amount proportional to its respective congestion. Henge  uses  the capacity  metric (Section~\ref{congestion}, eq.~\ref{eq:3}) to discover all congested operators in this chosen topology, i.e., operator capacity $>0.3$. It allocates each congested operator an extra number of threads based on the following equation:  
 
\vspace{-0.6cm}
	\begin{center}
		\begin{equation} 
		\small
 \left(\frac{Current\ Operator\ Capacity}{Capacity\ Threshold}-1
 \right)  \times 10
		\label{eq:executors}
		\end{equation}
	\end{center}
	\normalsize


Henge deploys this configuration change to a single topology on the cluster, and waits for the measured utilities to quiesce (this typically takes a minute or so in our configurations). No further actions are taken in the interim. It then measures the total cluster utility again, and if it improved, Henge  continues its operations in further rounds, in the Not Converged State. If this total utility reaches the maximum value (the sum of  maximum utilities of all topologies), then Henge cautiously continues monitoring the recently configured topologies for a while (4 subsequent rounds in our setting). If they all  stabilize, Henge moves the cluster to the Converged state. 

A topology may improve only marginally after being given more resources in a reconfiguration, e.g., utility increases $<5$\%. In such a case, Henge retains the reconfiguration change but skips this particular topology in the near future rounds. This is because the topology may have plateaued in terms of marginal benefit from getting more threads. Since the cluster is dynamic, this black-listing of a topology is not permanent but is allowed to expire after a while (1 hour in our settings), after which the topology will again be a candidate for reconfiguration.

As reconfigurations are exploratory steps in the  state space search, total system utility may  decrease after a step. Henge employs two actions called {Reduction} and {Reversion} to handle such cases.

\vspace{-0.1cm}

\subsubsection{Reduction}
\label{reduction}

If a Reconfiguration causes total system utility to drop, the next action is either a Reduction or a Reversion. Henge performs Reduction if and only if three conditions are  true: (a) the  cluster is congested (we detail below what this means),  (b) there is at least one SLO-satisfying topology, and (c) there is no past history of a Reduction action. 


First, CPU load is defined as the number of processes that are running or runnable on a machine~\cite{load}. A machine's load should be $\leq$ number of available cores, ensuring maximum utilization and no over-subscription.  
As a result, Henge considers a machine congested if its CPU load exceeds its number of cores. Henge considers a {\it cluster congested} when it has a majority of its machines congested. 

If a Reconfiguration drops utility and results in a  congested cluster, Henge executes  Reduction to reduce congestion. For all topologies {\it meeting} their SLOs, it finds all their un-congested operators (except spouts) and reduces their parallelism level by  a large amount (80\% in our settings). If this results in SLO misses, such topologies will be considered in future reconfiguration rounds. To minimize intrusion, Henge limits  Reduction to once per topology; this is reset if external factors  change (input rate, set of jobs, etc.).

Akin to backoff mechanisms~\cite{jacobson1988congestion}, massive reduction is the only way to free up a lot of resources at once,  
so that future reconfigurations may have a positive effect. Reducing threads also decreases their context switching overhead.


Right after a reduction, if the next reconfiguration drops cluster utility again while keeping the cluster congested (measured using CPU load), Henge recognizes that performing another reduction would be futile.  This is a typical ``lockout" case, and  Henge resolves it by performing Reversion. 

 \vspace{-0.1cm}
\subsubsection{Reversion}
\label{reversion}

If a Reconfiguration drops utility and a Reduction is not possible (meaning that at least one of the conditions (a)-(c) in Section~\ref{reduction} is not true), Henge performs Reversion.

Henge sorts through its history of Reconfigurations and picks the one that maximized system utility. It moves the system back to this past configuration by resetting the resource allocations of all jobs to values in this past configuration and moves to the Converged state. Here, Henge essentially concludes that it is impossible to further optimize cluster utility, given this workload. Henge maintains this configuration until changes like further SLO violations occur, which necessitate reconfigurations.





If a large enough drop ($>5 \%$) in utility occurs in this Converged state (e.g., due to new jobs, or input rate changes), Henge infers that as reconfigurations cannot be a cause of this drop, the workload of topologies must have changed. As all past actions no longer apply to this change in behavior, Henge  forgets all history of past actions and moves to the Not Converged state. This means that in future reversions, forgotten states will not be available. This reset allows Henge to start its state space search afresh. 

\subsection{Discussion}
\noindent{\bf Online vs. Offline State Space Search:} 
Henge prefers an online state space search. In fact, our early attempt at designing Henge was to perform offline state space exploration (e.g., through simulated annealing), by measuring SLO metrics (latency, throughput) and using analytical models to predict their relation to resources allocated to the job.

\begin{figure}[ht]
	\vspace{-0.3cm}	
	\begin{center}
		\includegraphics[width=1.0\columnwidth]{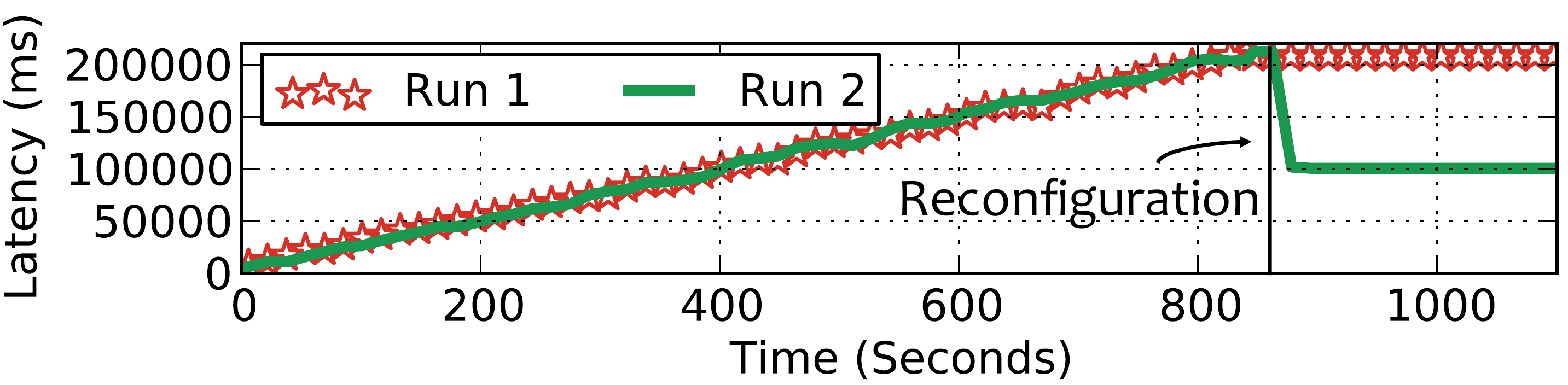}
		\caption{{\bf Unpredictability in Modern Stream Processing Engines:} \it Two runs of the same topology (on 10 machines) being given the same extra computational resources (28 threads, i.e., executors) at 910 s, react differently.}%
		\vspace{-0.4cm}	
		\label{same-step}
	\end{center}

\end{figure}

The offline approach turned out to be impractical. Analysis and prediction is complex and often turns out to be inaccurate  for stream processing systems, which are very dynamic in nature. (This phenomenon has also been observed in other distributed scheduling domains, e.g., see ~\cite{automatic-tuning,Morpheus,ousterhout2017monotasks}.)  
We show an example in Fig.~\ref{same-step}. The figure shows two runs of the same Storm job on 10 machines. In both runs we gave the job equal additional thread resources (28 threads) at t=910 s. Latency drops to a lower value in run 2, but only stabilizes in run 1. 
This is due to differing CPU resource consumptions across the runs. More generally, we find that natural fluctuations occur commonly in an application's throughput and latency; left to itself an application's performance changes and degrades gradually over time. We observed this for all our  actions: reconfiguration, reduction, and reversion. Thus, we concluded that online state space exploration would be more practical.


\noindent{\bf Tail Latencies: } Henge can also admit SLOs expressed as tail latencies (e.g., 95th percentile, or 99th percentile). Utility functions are then expressed in terms of  tail latency and the state machine remains unchanged.


\noindent{\bf Statefulness, Memory Bottlenecks: } The common case among topologies is stateless operators that are CPU-bound, and our exposition so far is thus focused. Nevertheless, Henge gracefully handles stateful operators and memory-pressured nodes (evaluated in Sections~\ref{sec:stateful_expts}, \ref{sec:memory-utilization}).




\section{Juice: Definition and Algorithm}
\label{juice}

	As described in Section~\ref{introduction}, we wish to design a throughput metric (for use in throughput SLOs), in a way that is independent of input rate. Henge uses a new metric called {\it juice}. Juice defines what fraction of the input data is being processed by the topology per unit time. It lies in the interval [0, 1], and a value of 1.0 means all the input data that arrived in the last time unit has been processed. Thus, the user can set throughput requirements as a percentage of the input rate (Section~\ref{job-utilities}), and Henge subsequently attempts to maintain this even as input rates change.

Any algorithm that calculates juice should be: 

\noindent{\it 1. Code Independent:} It should be independent of the operators' code, and should be calculate-able by only considering the number of tuples generated by  operators.

\noindent{\it 2. Rate Independent:} It should be independent of input  rates. 


\noindent{\it 3. Topology Independent:} It should be independent of the shape and structure of the topology. 

\noindent{\bf Juice Intuition: } Overall, juice is formulated to reflect the {\it global} processing efficiency of a topology. 
 We define per-operator contribution to juice as the proportion of input passed in originally \emph{from the source} that the operator processed in a given time window. This reflects the impact of that operator \emph{and} its upstream operators, on this input. The juice of a topology is then the normalized sum of  juice values of all its sinks. 

\noindent{\bf Juice Calculation: }Henge calculates juice in configurable windows of time (unit time). We define {\it source input} as the tuples that arrive at the input operator in a unit of time. For each operator $o$ in a topology that has $n$ parents, we define $T_{o}^{i}$ as the total number of tuples sent out from its ${i}^{th}$ parent per time unit, and $E_{o}^{i}$ as the number of tuples that operator $o$ executed (per time unit), from those received from parent $i$.

The per-operator contribution to juice, $J_{o}^{s}$, is the proportion of source input {\it sent} from source $s$ that operator $o$ received and processed. 
Given that  $J_{i}^{s}$ is the juice of $o$'s $i^{th}$ parent, then $J_{o}^{s}$ is: 

\vspace{-1.0cm}
\small
\begin{center}
\begin{equation} \label{eq:1}
J_{o}^{s} = \sum\limits_{i=1}^n \left( J_{i}^{s} \times \frac{E_{o}^{i}}{T_{o}^{i}} \right)
\end{equation}
\end{center}
\vspace{-0.2cm}
\normalsize

A spout $s$ has no parents, and its juice: $J_{s}=\frac{E_{s}}{T_{s}} = 1.0$ .

In eq.~\ref{eq:1}, the fraction $\frac{E_{o}^{i}}{T_{o}^{i}}$ reflects the proportion of tuples an operator received from its parents, and processed successfully.  If no tuples waiting in queues, this fraction is equal to 1.0. By multiplying this value with the parent's juice we accumulate through the topology the effect of all upstream operators. 

We make two important observations. In the term $\frac{E_{o}^{i}}{T_{o}^{i}}$,  it is critical to take the denominator as the number of tuples {\it sent} by a parent rather than received at the operator. This allows juice: a)  to account for data splitting at the parent (fork in the DAG), and b) to be reduced by  tuples dropped by the network. The numerator is the number of {\it processed} tuples rather than the number of output tuples -- this allows juice to generalize to operator types whose processing  may drop tuples (e.g., filter).

Given all operator juice values, a topology's juice can be calculated by normalizing w.r.t. number of sources:
\vspace{-0.5cm}
\begin {center}
\begin {equation} \label{eq:2}
\frac{\sum\limits_{Sinks\ s_{i},\  Sources\ s_{j}}(J_{s_{i}}^{s_{j}} )}{ { Total \ Number \ of \ Sources}}
\end {equation}
\end {center}
\vspace{-0.1cm}


If no tuples are lost in the system, the numerator sum is equal to the number of sources. To ensure that juice stays below 1.0, we normalize the sum with the number of sources. 



\noindent {\bf  Example 1: }  Consider Fig.~\ref{simple-topology} in Section~\ref{background}. $J_{A}^{s} = 1 \times \frac{10K} {10K} = 1$ and $J_{B}^{s} = J_{A}^{s} \times \frac{8K} {16K} = 0.5$. 
B has a $T_{B}^{A}$ of 16K and not 8K, since B only receives half the tuples that were sent out by operator A, and its per-operator juice should be in context of only this half (and not all source input).

The value of $J_{B}^{s}=0.5$ indicates that B processed only half the tuples sent out by parent A. This occurred as the parent's output was split among children. (If (alternately) B and C were sinks (if D were absent from the topology), then their  juice values would sum up to the topology's juice.). D has two parents: B and C. C is only able to process 6K as it is congested. Thus, $J_{C}^{s}= J_{A}^{s} \times \frac{6K} {16K}=0.375$. $T_{D}^{C}$ thus becomes 6K. Hence, $J_{D}^{C} = 0.375 \times \frac{6K} {6K} = 0.375$. $J_{D}^{B}$ is simply $0.5 \times \frac{8K} {8K} = 0.5$. We sum the two and obtain $J_{D}^{s} = 0.375 + 0.5 = 0.875$. It is less than 1.0 as C was unable to process all tuples due to  congestion.\\ 

\noindent \textbf{Example 2 (Topology Juice with Split and Merge):} 

\begin{figure}[h!]
\vspace{-0.3em}
\includegraphics[width=1\columnwidth]{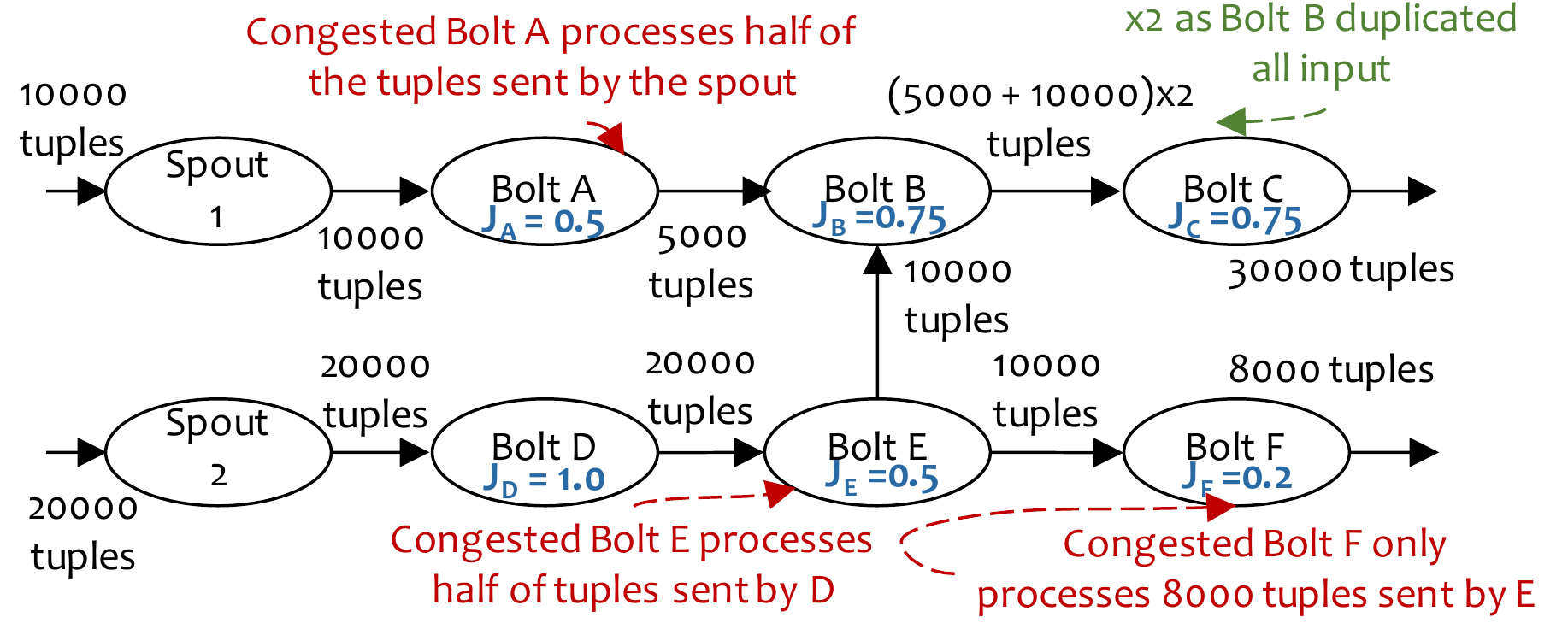}
\vspace{-1.5em}
\caption{{\bf {Juice Calculation in a Split and Merge Topology.}} }
\label{complicated-topology}
\vspace{-0.1cm}
\end{figure}

In Fig.~\ref{complicated-topology}, we show how our approach generalizes to: a) multiple sources (spout 1 \& 2), and b) operators splitting output (E to B and F) and c) operators with multiple input streams (A and E to B). Bolt A has a juice value of 0.5 as it can only process half the tuples spout 1 sent it. Bolt D has a juice value of 1.0. 50\% of the tuples from D to E are unprocessed due to congestion at E.  E passes its tuples on to B and F: both of them get half of the total tuples it sends out. Therefore, B has juice of 0.25 from E and 0.5 from A  ($0.25+0.5=0.75$). 20\% of the tuples E sent F are unprocessed at F as it is congested, so F has  a juice value of $0.25\times0.8=0.2$. C processes as many tuples as B sent it, so it has the same juice as B (0.75). The juice of the topology is the sum of the juices of the two sinks, normalized by the number of sources. Thus, the topology's juice is  $\frac{0.2+0.75}{2} = 0.475$.



\normalsize

\noindent{\bf Some Observations:} First, while our description used unit time, our implementation calculates juice using a sliding window of 1 minute, collecting data in sub-windows of length 10 s. This needs only loose time synchronization across nodes (which may cause juice values to momentarily exceed 1, but does not affect our logic). Second, eq.~\ref{eq:2} treats all processed tuples equally--instead, a weighted sum could  be used to capture the higher importance of some sinks (e.g., sinks feeding into a dashboard). Third, processing guarantees (exactly, at least, at most once) are orthogonal to the juice metric. Our experiments use the non-acked version of Storm (at most once semantics), but Henge also works with the acked version of Storm (at least once semantics).

\vspace{-0.1cm}
\section{Implementation}
\label{system-architecture}

We integrated Henge into Apache Storm~\cite{storm_web}. Henge involves 3800 lines of Java code. It is an implementation of the predefined IScheduler interface. The scheduler runs as part of the Storm Nimbus daemon, and  
is invoked by Nimbus periodically every 10 seconds. Further changes were made to Storm Config, allowing users to set topology SLOs and utility functions while writing topologies.

\begin{figure}[h!]
	\begin{center}
 	 \vspace{-0.3cm}
		\includegraphics[width=1\columnwidth]{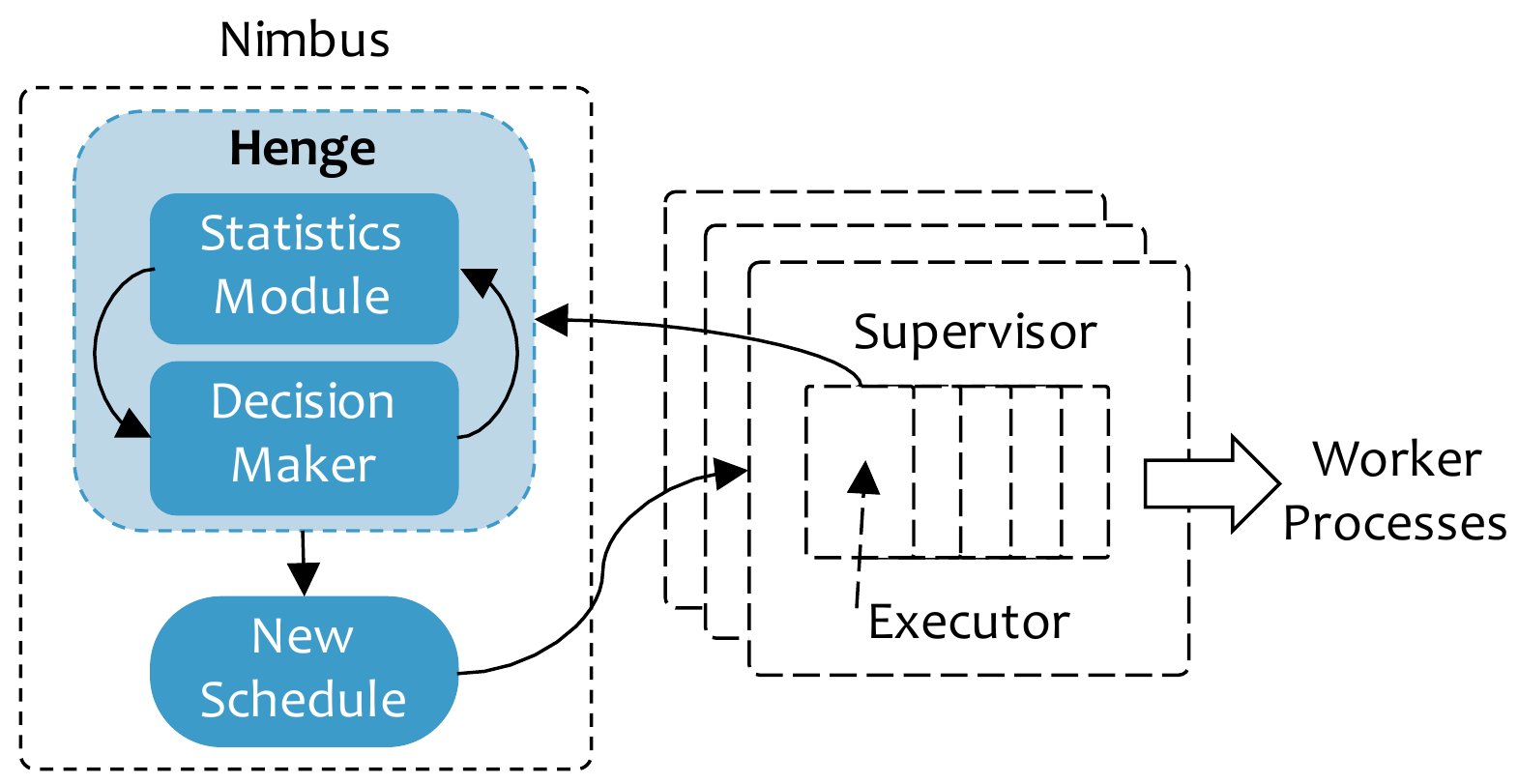}
	\end{center}
	\vspace{-0.5cm}
	\caption{{\bf Henge Implementation:} \it Architecture in Apache Storm.}
	\label{henge-architecture}
 	\vspace{-0.3cm}
\end{figure}

Henge's architecture is shown in Fig.~\ref{henge-architecture}. The Decision Maker implements the Henge state machine of Section \ref{policy}. The Statistics Module continuously calculates cluster and per-topology metrics such as the number of tuples processed by each task of an operator per topology, the end-to-end latency of tuples, and the CPU load per node. This information is used to produce useful metrics such as juice and utility, which are passed to the Decision Maker. The Decision Maker runs the state machine, and sends commands to Nimbus to implement actions. 
The Statistics Module also tracks past states so that reversion can be performed. 







\vspace{-0.1cm}
\section{Evaluation}
\label{evaluation}


We evaluate Henge with a variety of workloads, topologies, and SLOs. 



\noindent\textbf{Experimental Setup:} By default, our experiments used the Emulab cluster~\cite{White+:osdi02}, with machines (2.4 GHz, 12 GB RAM) running Ubuntu 12.04 LTS, connected via a 1 Gbps connection. Another machine runs Zookeeper~\cite{zookeeper_web} and Nimbus. Workers (Java processes running executors) are allotted to each of our 10  machines (we evaluate scalability later). 

\begin{figure}[h!]
	\centering
		\vspace{-0.2cm}
	\includegraphics[width=1.0\columnwidth]{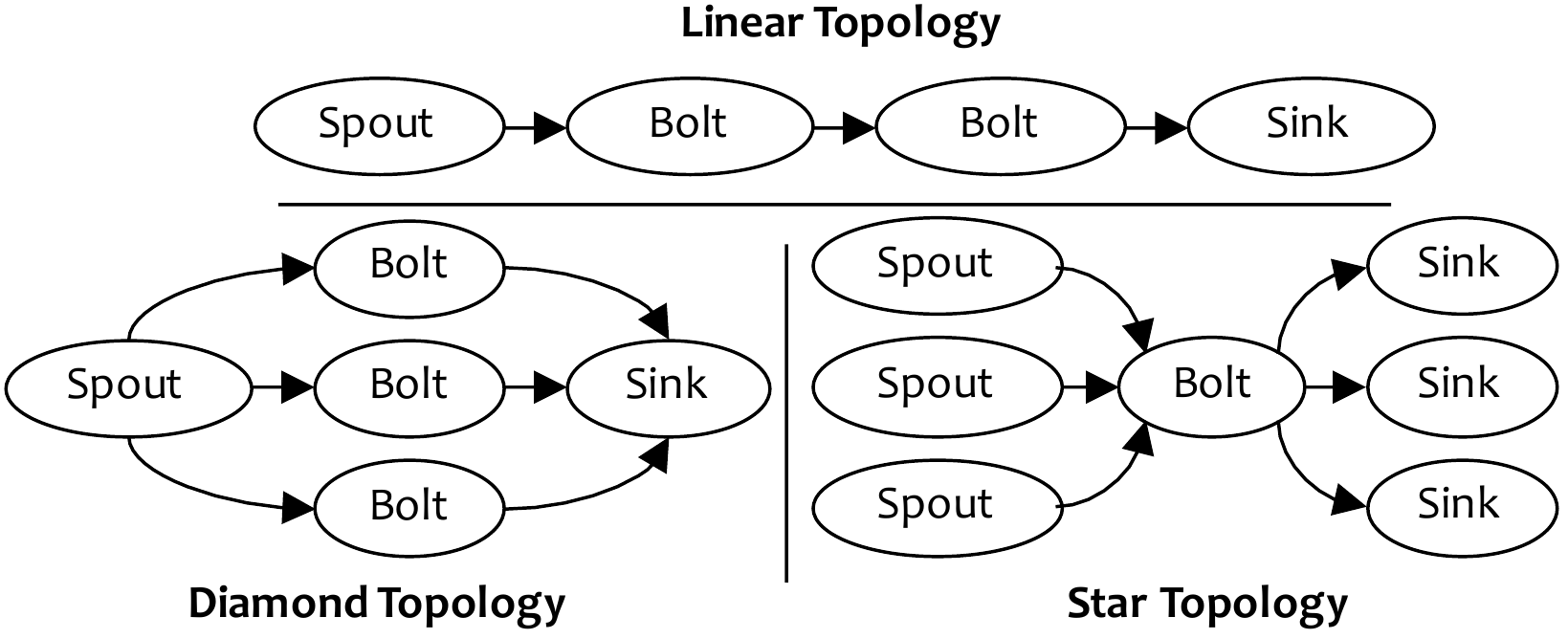}
	\vspace{-0.6cm}
	\caption{\bf Three Microbenchmark Topologies.}
	\label{topology-types}
	\vspace{-0.5cm}
\end{figure}
\vspace{-0.1cm}
\begin{figure}[h!]
	\centering
	\includegraphics[width=1.0\columnwidth]{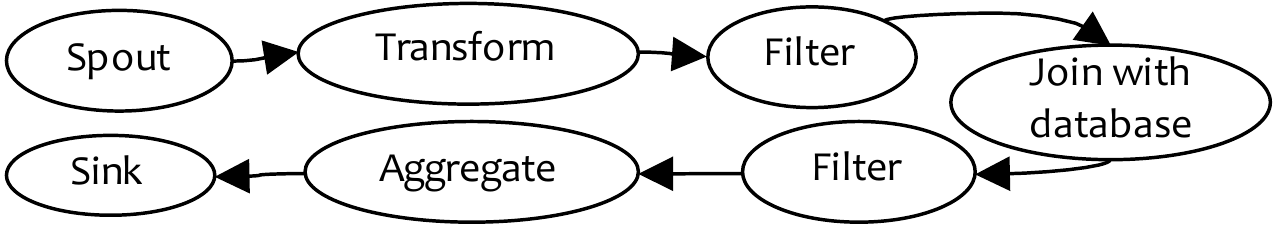}
	\vspace{-0.6cm}
	\caption{\bf PageLoad Topology from Yahoo!.}
	\label{pageload}
	\vspace{-0.5cm}
\end{figure}

\noindent\textbf{Topologies: } For evaluation, we use both: a) micro-topologies that are possible sub-parts of larger topologies~\cite{xu:16:ic2e}, shown in Fig.~\ref{topology-types}; and b) a production topology from Yahoo! Inc.~\cite{xu:16:ic2e}--this topology is called ``PageLoad" (Fig.~\ref{pageload}).  Operators are the ones that are most commonly used in production: filtering, transformation, and aggregation. In each experimental run, we initially allow topologies to run for 900 s without interference (to stabilize and to observe their performance with vanilla Storm), and then enable Henge to take actions. All topology SLOs use the knee utility function of Section~\ref{job-utilities}. Hence, below we use ``SLO'' as a shorthand for the SLO threshold.

\subsection{Juice as a Performance Indicator} 
\label{juice-evaluation}

\begin{figure}[h]
		\begin{center}
			\includegraphics[width=1.0\columnwidth]{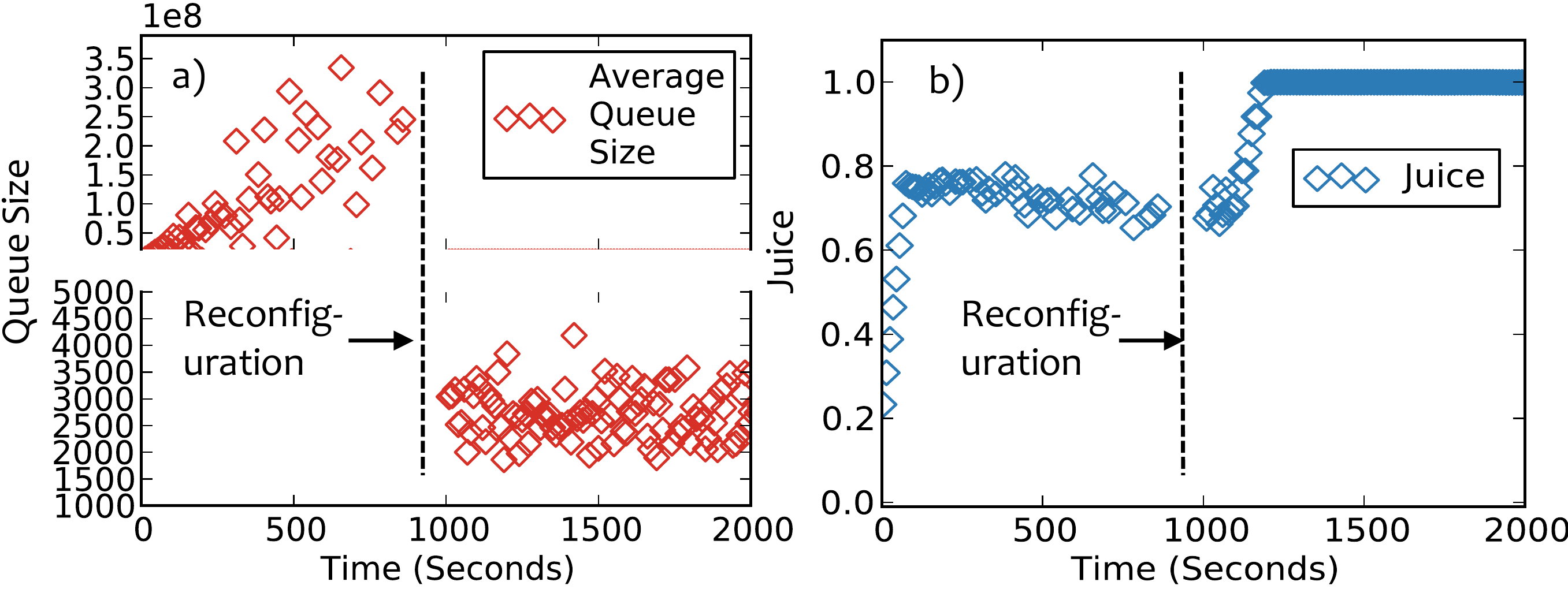}
		\end{center}
	\vspace{-0.5cm}
	\caption{{\bf Juice vs. Queue Size:} \it Inverse Relationship.}
    \vspace{-0.1cm}
	\label{queue-size}
\end{figure}

 \noindent{\bf Juice is an indicator of queue size: } Fig.~\ref{queue-size} shows the inverse correlation between topology juice and queue size at the most congested  operator of a PageLoad topology. Queues buffer incoming data for operator executors, and longer queues imply slower execution rate and higher latencies. Initially queue lengths are high and erratic--juice captures this by staying well below 1. At the reconfiguration point (910 s) the operator is given more executors, and juice converges to 1 as queue lengths fall, stabilizing by 1000 s. 
 
\vspace{-0.3cm}



\begin{figure}
	\begin{center}
		\vspace{-0.1cm}
\includegraphics[width=1.0\columnwidth]{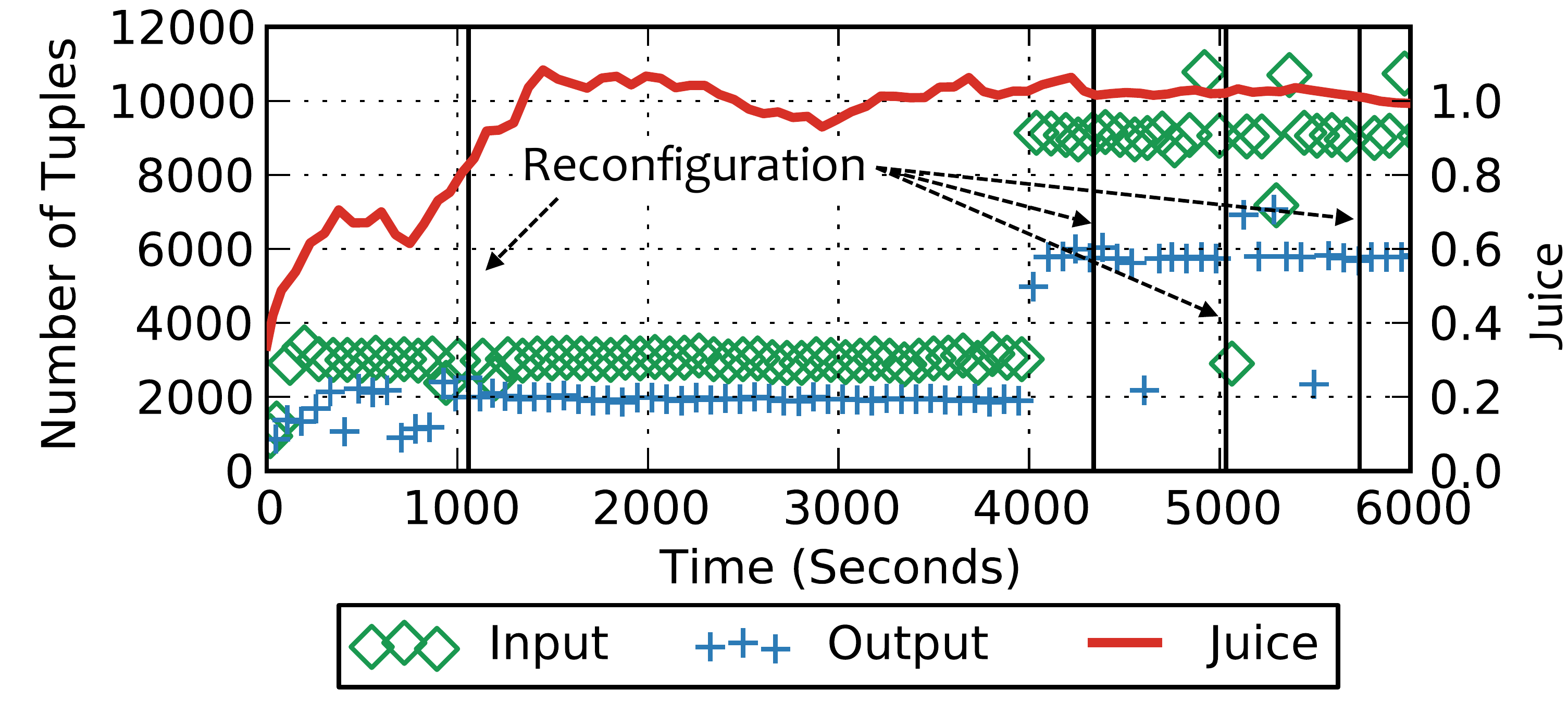}\vspace{-0.2cm}
\caption{{\bf Juice is Rate-Independent:} \it Input rate is increased by 3 $\times$ at 4000 s, but juice does not change. When juice falls  to 0.992 at 4338 s, Henge  stabilizes it to 1.0 by 5734 s. }
\vspace{-0.5cm}
\label {page-load-tuples} 
\end{center}
\end{figure}

\noindent{\bf Juice is independent of operations and input rate: } In Fig.~\ref{page-load-tuples}, we run 5 PageLoad topologies on one cluster, and show data for one of them. Initially juice stabilizes to around 1.0, near t=1000 s (values above 1 are due to synchronization errors, but they don't affect our logic). PageLoad filters tuples, thus output rate is $<$ input rate--however, juice is 1.0 as it shows that all input tuples are being processed.

Then at 4000 s, we triple the input rate to all tenant topologies. Notice that juice stays 1.0. Due to natural fluctuations, at 4338 s, PageLoad's juice drops to 0.992. This triggers  reconfigurations (vertical lines) from Henge, stabilizing the system by 5734 s, maximizing cluster utility.


\subsection{Henge Policy and Scheduling} 
\label{policy-evaluation}

\subsubsection{Impact of Initial Configuration}
\begin{figure}[h!]
	\begin{center}
	\vspace{-0.2cm}
\includegraphics[width=1.0\columnwidth]{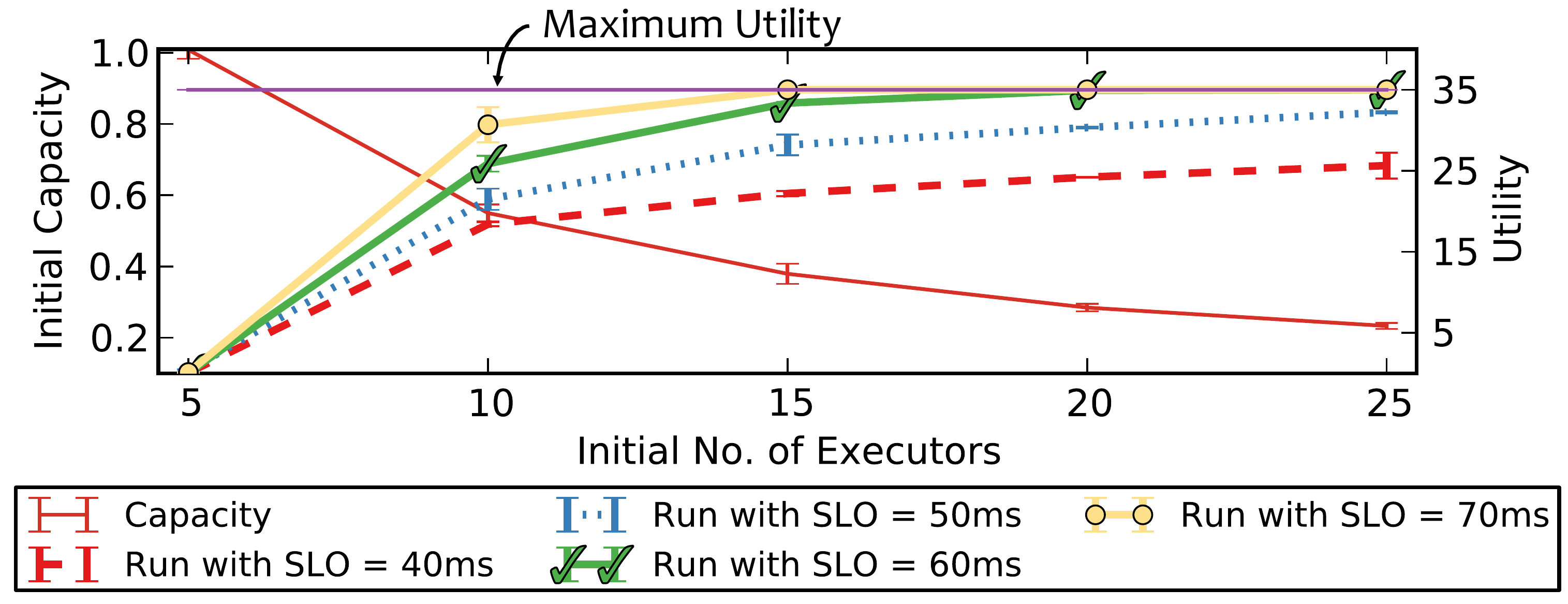}
\vspace{-0.3cm}
\caption{{\bf Performance vs. Resources in Apache Storm:} \it The x-axis shows initial parallelism of one intermediate operator in a linear topology. Left y-axis shows initial capacity of the operator. Right y-axis shows stable utility  reached without using Henge.} 
		\label{changing-capacity}
	\end{center}
\vspace{-0.5cm}
\end{figure}

\begin{figure}[t]
\vspace{-0.8cm}
        \centering
        \includegraphics[width=1\columnwidth]{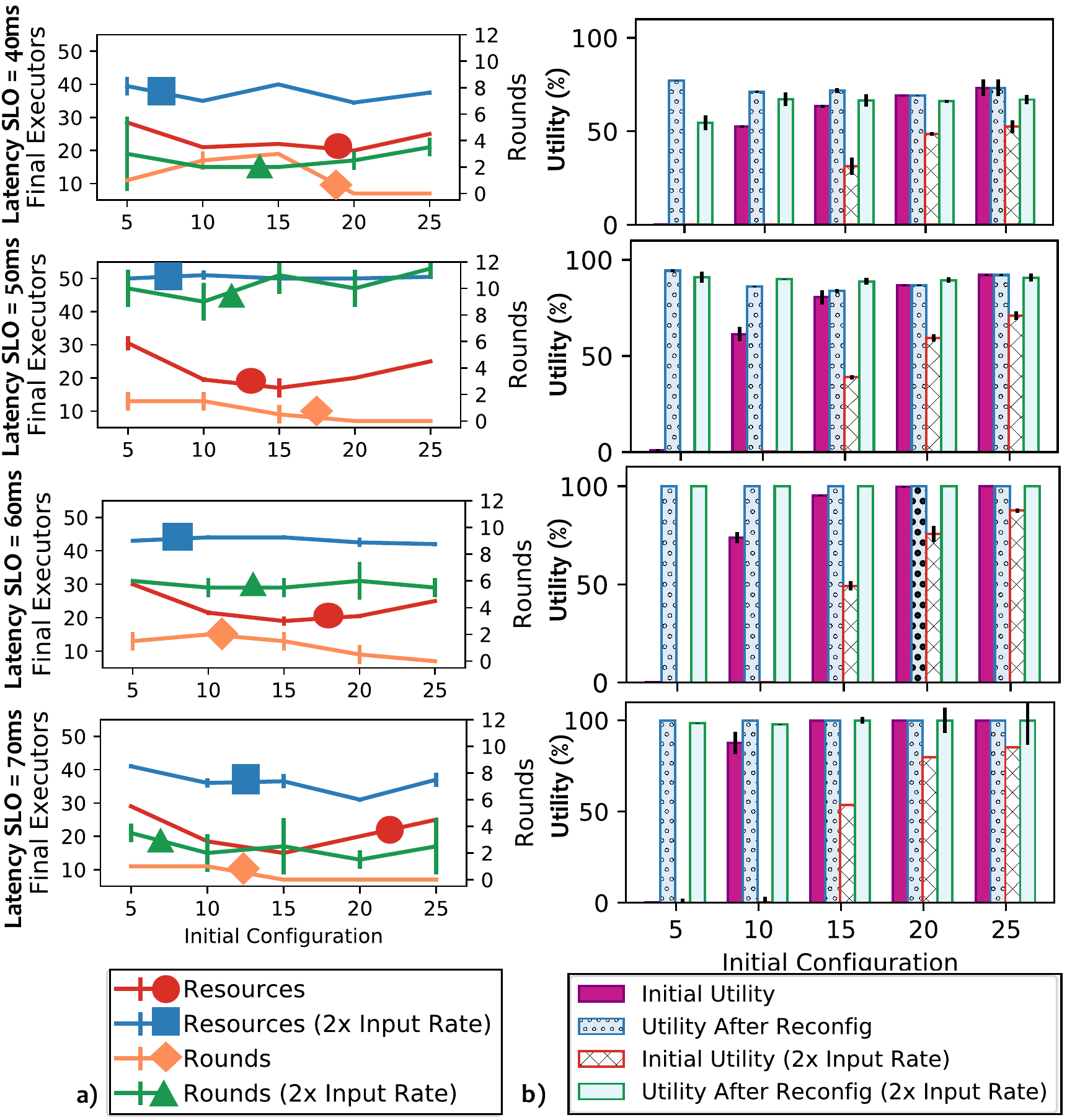}
        \vspace{-0.5cm}
\caption{{\bf Effect of Henge on Figure~\ref{changing-capacity}'s initial configurations:} 
\it SLOs become more stringent from bottom to top. We also explore a 2  $\times$ higher input rate. a) Left y-axis shows final parallelism level Henge assigned to each operator. Right y-axis shows number of rounds required to reach said parallelism level. b) Utility values achieved before and after Henge. }
    \label{changing-congestion}
\vspace{-0.4cm}
\end{figure}

\noindent{\bf State Space: }  Fig.~\ref{changing-capacity} illustrates the state space that Henge needs to navigate. These are runs {\it without} involving Henge. We vary the initial number of executors for an intermediate operator. Fewer initial executors   (5, 10) lead to a high capacity (indicating congestion: Section~\ref{congestion}) and consequently the topology is  unable to achieve its SLO. From the plot, the more stringent the SLO, the greater the number of executors needed to reach max utility. Except very stringent jobs SLOs (40, 50 ms) all others can meet their SLO. 


\noindent{\bf Henge In Action: } Now, we put Henge into action on  Fig.~\ref{changing-capacity}'s topology and initial state, with max utility 35. Fig.~\ref{changing-congestion} shows the effect of varying: a) initial number of executors (5 to 25), b) latency SLO (40 ms to 60 ms), and  c)  input rate. We plot converged utility, rounds needed, and executors assigned. 
 
 We observe that generally, Henge gives more resources to topologies with more stringent  SLOs and  higher  input rates. For instance, for a congested operator initially assigned 10 executors in a 70 ms SLO topology, Henge  reconfigures it to have an average of 18 executors, all in a single round. On the other hand, for a stricter 60 ms SLO it assigns 21 executors in two rounds. 
 When we double the input rate of these two topologies, the former is assigned 36 executors in two rounds and the latter is assigned 44, in 5 rounds.  



Henge convergence is fast. In Fig.~\ref{changing-congestion}a, convergence occurs within 2 rounds for a topology with a 60 ms  SLO. Convergence time increases for stringent SLOs and higher input rates. With the 2 $\times$ higher input rate convergence time is 12 rounds for stringent SLOs of 50 ms, vs. 7 rounds for 60 ms.

Henge always reaches  max utility (Fig.~\ref{changing-congestion}b) unless the SLO is unachievable (40, 50 ms SLOs). 
Since Henge aims to be minimally invasive, we do not explore operator migration (but we could use them orthogonally~\cite{twochoices,morethantwochoices,network-aware-placement}).

 With an SLO of 40 ms, Henge actually performs fewer reconfigurations and allocates less resources than with a laxer SLO of 50 ms. This is because the 40 ms topology gets black-listed earlier than the 50 ms topology ( Section~\ref{reversion}: recall this occurs if utility improves $<$ 5\% in a round). 

Overall, by black-listing  topologies with overly stringent SLOs and satisfying other topologies, Henge meets its goal of preventing resource hogging  (Section~\ref{summary}).



\subsubsection{Meeting SLOs}



\begin{figure}[h!]
	\vspace{-0.1cm}	
	\begin{center}
\includegraphics[width=1.0\columnwidth]{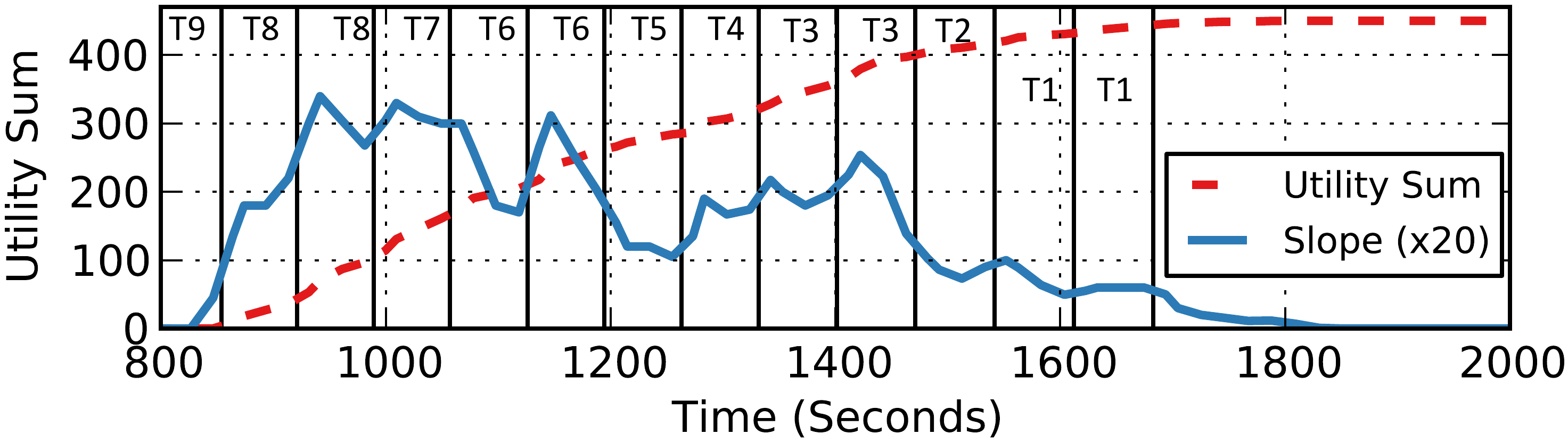} 
\vspace{-0.7cm}
\caption{{\bf Maximizing Cluster Utility:} \emph{Red (dotted) line is total system utility. Blue (solid) line is magnified slope of the red line. Vertical lines are reconfigurations annotated by the job touched. Henge reconfigures higher max-utility jobs first, leading to faster increase in system utility.}} %
		\label{discussion:mixed-topologies}
	\end{center}
\vspace{-0.25cm}
\end{figure}

\noindent{\bf Maximizing Cluster Utility: } To maximize  total cluster utility, Henge greedily prefers to reconfigure those topologies first which have a higher max achievable utility (among those missing their SLOs). In Fig.~\ref{discussion:mixed-topologies}, we run 9 PageLoad topologies on a cluster, with max utility values ranging from 10 to 90 in steps of 10. The SLO threshold for all topologies is 60 ms.  
Henge first picks T9 (highest max utility of 90),  leading to a sharp increase in total cluster utility at 950 s. Thereafter, it continues in this greedy way.  We observe some latent peaks when topologies reconfigured in the past stabilize to their max utility. For instance, at 1425 s we observe a sharp increase in the slope (solid) line as T4 (reconfigured at 1331 s) reaches its SLO threshold.  All topologies  meet their SLO within 15 minutes (900 s to 1800 s). 

\noindent{\bf Hybrid SLOs:} We evaluate a hybrid SLO that has separate thresholds for latency and juice, and two corresponding utility functions (Section~\ref{job-utilities}) with identical max utility values. The job's utility is then the average of these two utilities. 

\begin{figure}[h!]
	\begin{center}
	\vspace{-0.2cm}
\includegraphics[width=1.0\columnwidth]{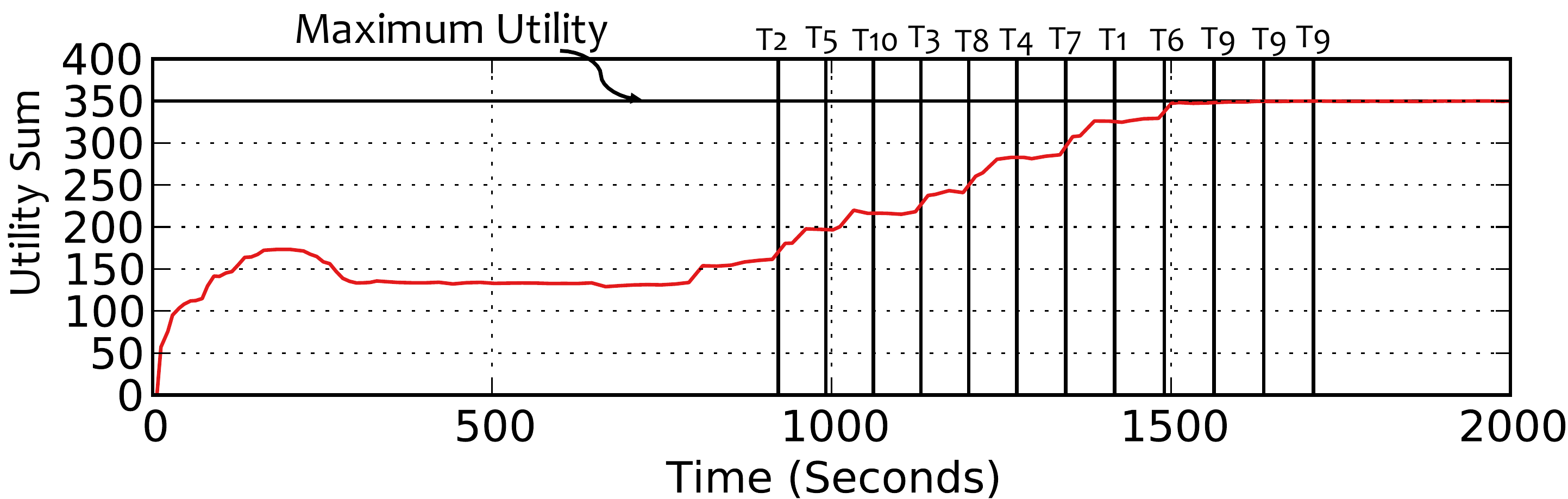} 
\vspace{-0.7cm}
\caption{{\bf Hybrid SLOs: } \it Henge Reconfiguration.}
		\label{cluster-intrusive}
	\end{center}
\vspace{-0.5cm}
\end{figure}

Fig.~\ref{cluster-intrusive} shows 10 (identical) PageLoad topologies with hybrid SLOs running on a cluster of 10 machines. Each topology has SLO thresholds of: juice 1.0, and latency 70 ms.  The max utility value of each topology is 35. Henge only takes about 13 minutes (t=920 s to t=1710 s) to reconfigure all topologies successfully to meet their SLOs.  9 out of 10 topologies required a single reconfiguration, and one (T9) required 3 reconfigurations. 

\subsubsection{Handling Dynamic Workloads}



\begin{figure}[h]
\vspace{-0.1cm}

                \centering
                \includegraphics[width=1\columnwidth]{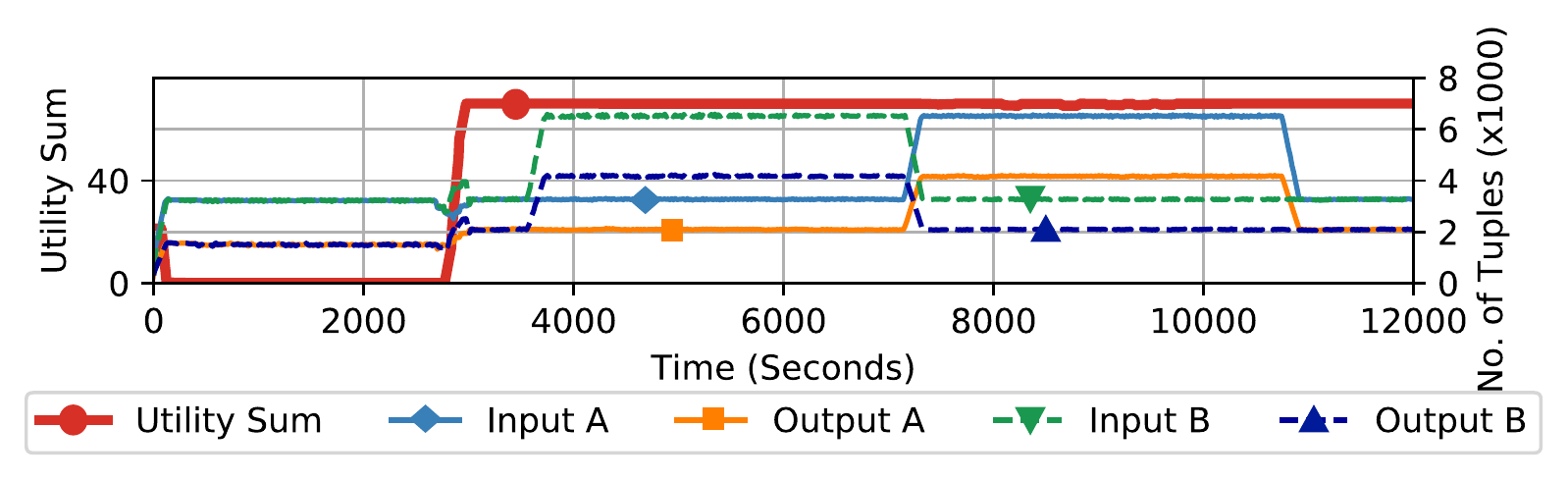}\vspace{-0.2cm}
        \caption{{\bf Spikes in Workload:} \it Left y-axis shows total cluster utility (max possible is $35\times 2=70$). Right y-axis shows the variation in workload as time progresses.}\vspace{-0.1cm}
		\label{discussion:step-workload}
	
\vspace{-0.3cm}
\end{figure}

\noindent{\bf A. Spikes in Workload:} 
Fig.~\ref{discussion:step-workload} shows the effect of a workload spike in Henge. Two different PageLoad  topologies A and B are subjected to input spikes. B's  workload spikes by 2 $\times$, starting from 3600 s. The spike lasts until  7200 s when A's spike (also 2 $\times$) begins.   Each topology's SLO is 80 ms with max utility is 35. Fig.~\ref{discussion:step-workload} shows that: i)  output rates keep up for both topologies both during and after the spikes, and ii) the utility stays maxed-out  during the spikes. In effect, Henge completely hides the effect of the input rate spike from the user. 

	


	\begin{figure}[h!]
	\begin{center}
\includegraphics[width=1.0\columnwidth]{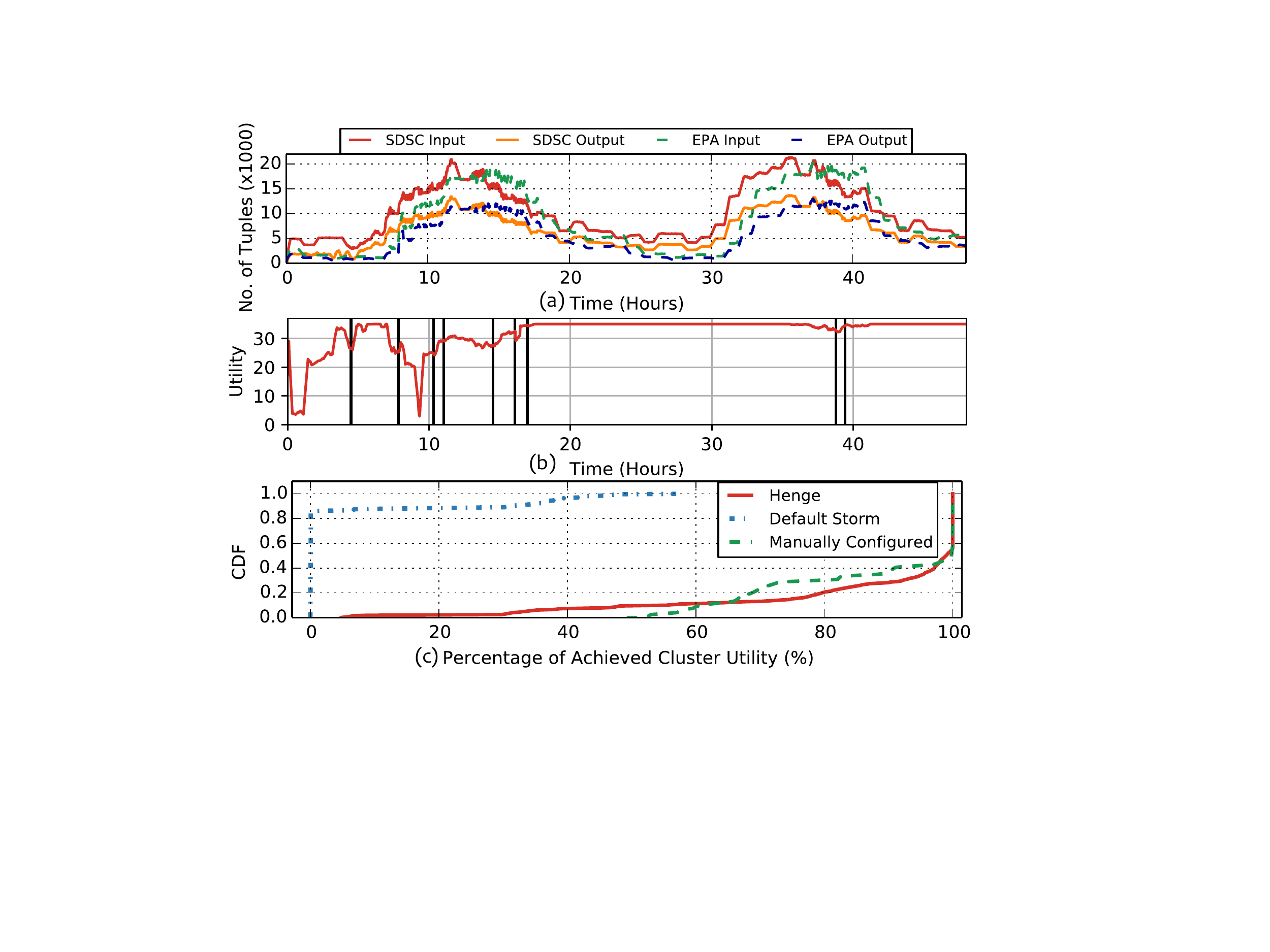}
\vspace{-0.8cm}


\caption{{\bf Diurnal Workload:} \it a) Input and output rates  over time, for two different diurnal workloads. b) Utility of a topology (reconfigured by Henge at runtime) with the EPA workload, c) CDF of SLO satisfaction for Henge, default Storm, and manually configured. Henge adapts during the first cycle and fewer reconfigurations are required thereafter. }

		\label{diurnal}
	\end{center}
\vspace{-0.8cm}
\end{figure}

\noindent{\bf B. Diurnal Workloads:} Diurnal workloads are common for stream processing in production, e.g., in e-commerce websites~\cite{decandia2007dynamo} and social media~\cite{naaman2012study}.  
We generated a  diurnal workload based on the distribution of the SDSC-HTTP ~\cite{web-trace1} and EPA-HTTP traces~\cite{web-trace2}, injecting them into PageLoad topologies.  
5 topologies are run with the SDSC-HTTP trace and concurrently, 5 other topologies are run with the EPA-HTTP trace. All 10 topologies have max-utility=10 (max achievable cluster utility=350), and a latency SLO of 60 ms.  

Fig.~\ref{diurnal} shows the results of running 48 hours of the trace (each hour mapped to 10 min intervals). In Fig.~\ref{diurnal}a  workloads increase from hour 7 of the day, reach their peak by  hour 13$\frac{1}{3}$, and then start to fall. Within the first half of Day 1, Henge successfully reconfigures  all 10 topologies, reaching by hour 15 a cluster utility that is 89\% of the max. 

Fig.~\ref{diurnal}b shows a topology running  the EPA workload (other topologies exhibited the same behavior). Observe how Henge reconfigurations from hour 8 to 16  adapt to the fast changing workload. This also results in fewer SLO violations during the second peak  (hours 32 to 40). Thus, even without adding resources, Henge tackles diurnal workloads. 


Fig.~\ref{diurnal}c shows the CDF of SLO satisfation for the three systems. Default Storm performs  poorly, giving 0.0006\% SLO satisfaction at the median, and 30.9\% at the 90th percentile. ({This means that 90\% of the time, default Storm provided a total of at most 30.9\% of the cluster's max achievable utility.}) Henge yields 74.9\% , 99.1\%,  and 100\% SLO satisfaction at the 15th, 50th, and 90th percentiles respectively. 

Henge is also better than manual configurations. We manually configured all topologies to meet their SLOs at  median load. 
These provide 66.5\%, 99.8\% and 100\% SLO satisfaction at the 15th, 50th and 90th percentiles respectively. Henge is better than manual configurations from the 15th to 45th percentile, and comparable from then onwards. 

Henge has an average of 88.12\% SLO satisfaction rate, whereas default Storm and manually configured topologies provide an average of  4.56\% and 87.77\% respectively. Thus, Henge provides 19.3 $\times$  better SLO satisfaction than default Storm, and performs better than manual configuration.








\subsection{Production Workloads}  

We configured the sizes of 5  PageLoad topologies based on data from a Yahoo! Storm production cluster and Twitter datasets~\cite{apollo-datasets}, shown in Table~\ref{table:yahoo-job-dist}. We use 20 nodes each with 14 worker processes. For each topology, we inject an input rate proportional to its number of workers. T1-T4 run sentiment analysis on Twitter workloads from 2011~\cite{apollo-datasets}. T5 processes logs at a constant rate. Each topology has a latency SLO threshold of 60 ms and max utility value of 35. 

\begin{table}[ht]
\begin{center}
\small
 \begin{tabular}{c c c c} 
 \toprule
Job & Workload & Workers & Tasks  \\ \hline
 T1 & Analysis (Egypt Unrest) & 234 & 1729 \\ \hline
 T2 & Analysis (London Riots)  & 31 & 459 \\\hline
 T3 &  Analysis (Tsunami in Japan)  & 8 & 100 \\\hline
 T4 &  Analysis (Hurricane Irene) & 2 & 34 \\\hline
 T5 &  Processing Topology & 1 & 18  \\ \bottomrule
\end{tabular}
\end{center}
\vspace{-0.5cm}
\caption{{\bf Job and Workload Distributions in Experiments: } \it Derived from Yahoo! production clusters, using Twitter Datatsets for T1-T4. (Experiments in Figure~\ref{prod}.)}
\label{table:yahoo-job-dist}
\vspace{-0.3cm}
\end{table}
\normalsize

\begin{figure}[h!]
\vspace{-0.2cm}	
	\begin{subfigure}{0.45\textwidth}
                \centering
                \includegraphics[width=1.0\columnwidth]{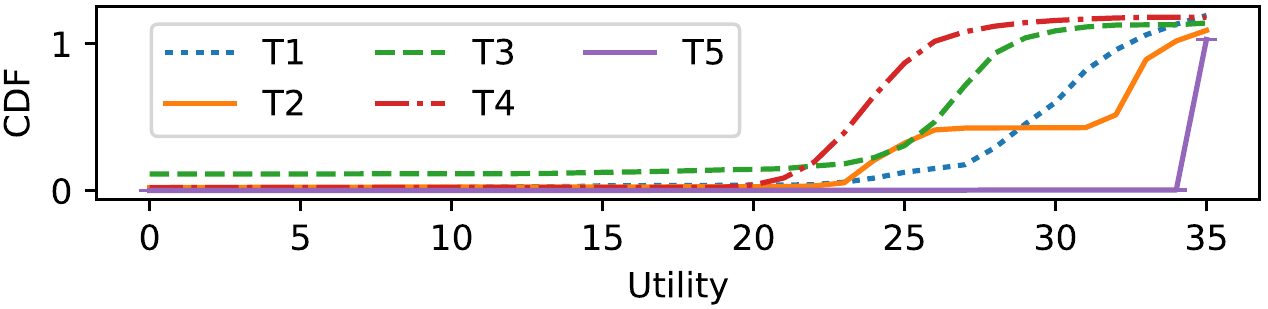}\vspace{-0.3cm}
        \caption{\it CDF of the fraction of total time  tenant topologies provide a given SLO satisfaction. Max utility for each topology is 35.}
            \label{prod-a}
        \end{subfigure}
        \begin{subfigure}{0.45\textwidth}
                \centering
                \includegraphics[width=1.0\columnwidth]{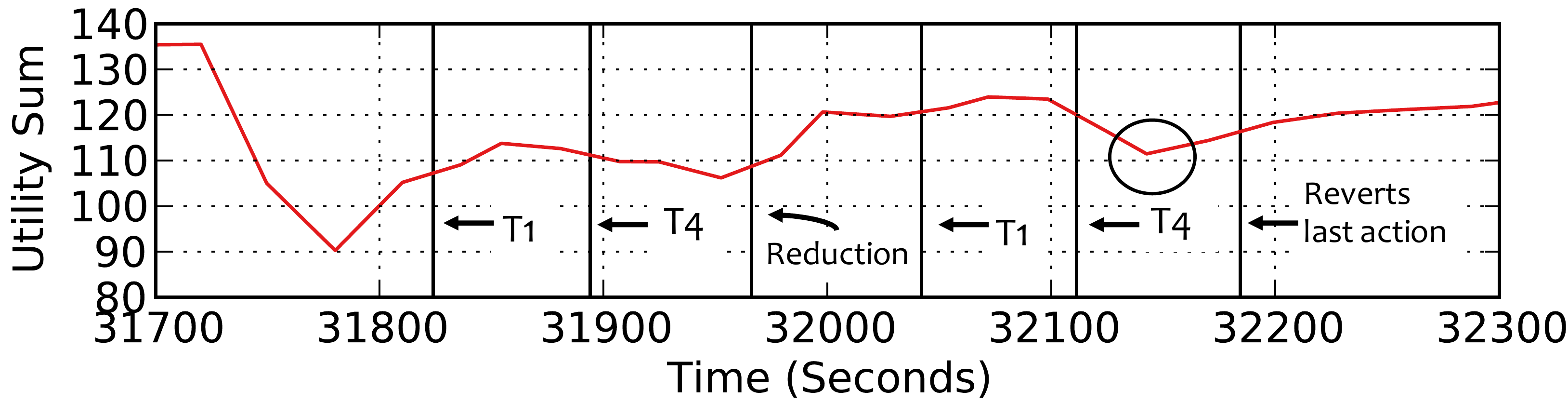}\vspace{-0.1cm}
        \caption{\it Reconfiguration at 32111 s causes a drop in total system utility. Henge reverts the configuration of all tenants to that of 32042 s. Vertical lines show Henge actions for particular jobs.}\vspace{-0.2cm}
            \label{prod-b}
        \end{subfigure}\vspace{-0.05cm}
    		\caption{\bf Henge on Production Workloads.}
		\label{prod}
 \vspace{-0.1cm}
\end{figure}

This is an extremely  constrained cluster where not all SLOs can be met. Yet, Henge maximizes cluster utility. Fig.~\ref{prod-a} shows the CDF of the fraction of time each topology provided a given utility (including the initial 900 s where Henge is held back). T5 shows the most improvement (at the 5th percentile, it has 100\% SLO satisfaction), whereas T4 shows the worst performance (at the median, its utility is 24, which is 68.57\% of 35). 
The median SLO satisfaction for T1-T3 ranges from 27.0 to 32.3  (77.3\% and 92.2\% respectively). 

\noindent{\bf Reversion: } Fig.~\ref{prod-b} depicts Henge's reversion. At 31710 s, the system utility drops due to natural system fluctuations. This forces Henge to perform reconfigurations for two topologies (T1, T4). Since  system utility continues to drop, Henge is forced to reduce a topology (T5, which satisfies its SLO before and after reduction. As utility improves at 32042 s, Henge proceeds to reconfigure other topologies. However, the last reconfiguration causes another  drop in utility (at 32150 s). Henge reverts to the configuration that had the highest utility (at 32090 s). After this point, total cluster utility stabilizes at 120 (68.6\% of max utility). Thus, even under scenarios where Henge is unable to reach the max system utility it behaves gracefully, does not thrash, and converges quickly.

\subsubsection{Reacting to Natural Fluctuations}

\begin{figure}[h!] 
	\vspace{-0.4cm}
	\begin{center}
\includegraphics[width=1.0\columnwidth]{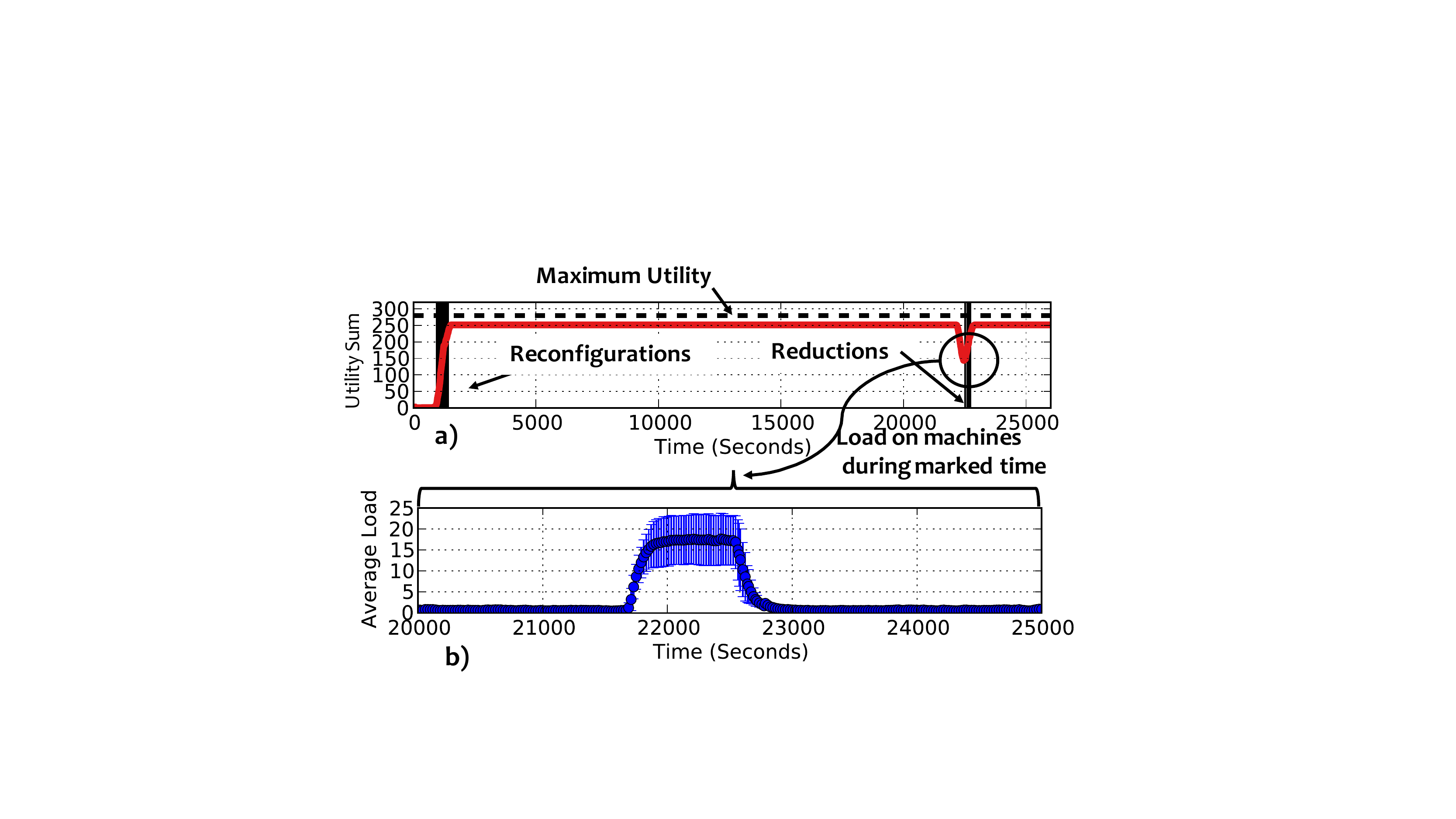}\vspace{-0.3cm}
\caption{{\bf Handling CPU Load Spikes:} \it a) Total cluster utility. 
b) Average CPU load on machines in the CPU spike interval.}
		\label{reduction-plot}
	\end{center}
\vspace{-0.5cm}
\end{figure}

Natural fluctuations occur in the cluster due to  load variation that arises from interfering processes, disk IO, page swaps, etc. Fig.~\ref{reduction-plot} shows such a scenario. We run 8 PageLoad topologies, 7 of which have an SLO of 70 ms, and the 8th SLO is 10 ms. Henge resolves congestion initially and stabilizes the cluster by 1000 s. At 21800 s, CPU load increases sharply due to OS behaviors (beyond Henge's control). Seeing the significant drop in cluster utility, Henge reduces two of the topologies (from among those meeting their SLOs). This allows other topologies to recover within 20 minutes (by 23000 s). Henge converges the system to the same total utility as before the CPU spike.

\subsection{Stateful Topologies} 
\label{sec:stateful_expts}



\begin{table}[ht]
\begin{center}
\vspace{-0.3cm}
 \small
 \begin{tabular}{ccc} 
 \toprule
Job Type & Avg. Reconfig.   & Average  Convergence \\ 
 & Rounds (Stdev) & Time (Stdev) \\ \hline
 Stateful & 5.5 (0.58) &  1358.7355 (58.1s) \\ \hline
 Stateless & 4 (0.82) &  1134.2235 (210.5s)  \\\bottomrule
\end{tabular}
\end{center}
\vspace{-0.3cm}
\caption{{\bf Stateful Topologies:} \it Convergence Rounds and Times for a cluster with Stateful and Stateless Topologies.} 
\label{table:stateful-properties}
\end{table}
\vspace{-0.2cm}
\normalsize

Henge handles stateful topologies gracefully,  alongside stateless ones. We ran four WordCount topologies with identical workload and configuration as T2 in Table~\ref{table:yahoo-job-dist}. Two of these topologies periodically checkpoint state to Redis (making them stateful) and have 240 ms latency SLOs. The other two topologies do not persist state in an external store and  have lower SLOs of  60 ms. Initially, none of the four meet their SLOs. Table~\ref{table:stateful-properties} shows results after convergence. Stateful topologies take 1.5 extra reconfigurations to converge to their SLO, and only 19.8\%  more reconfiguration time. This difference is due to external state checkpointing and recovery mechanisms, orthogonal to Henge.




\subsection{Scalability and Fault-tolerance}

\label{scale_ft}

We vary number of jobs and nodes, and inject failures.

\subsubsection{Scalability}

\begin{figure}[h!]  
\vspace{-0.2cm}
	\begin{center}
	\begin{subfigure}{0.5\textwidth}
                \includegraphics[width=1\columnwidth]{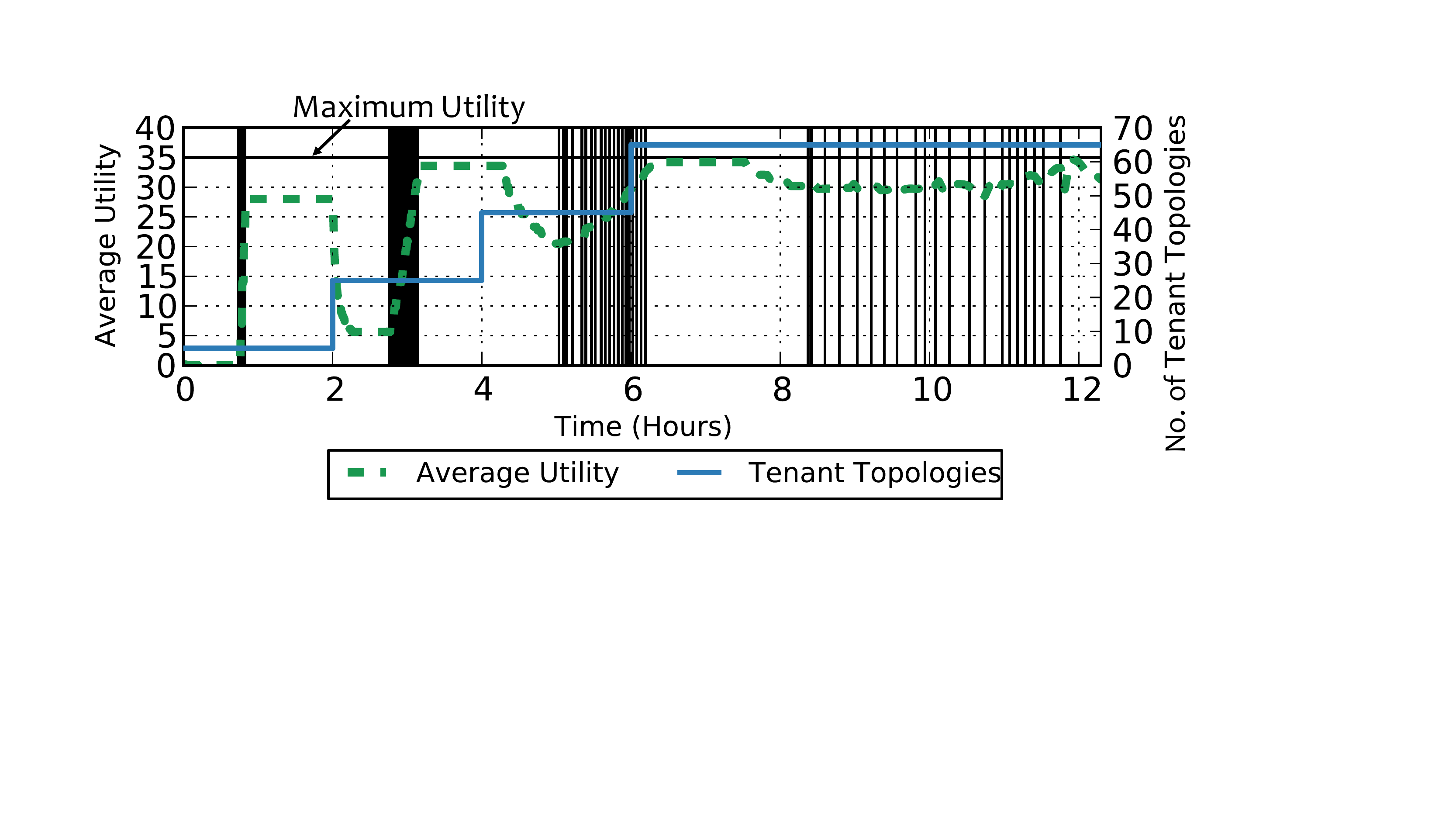}
        \caption{\it Green (dotted) line is average job utility. Blue (solid) line is number of job on  cluster. Vertical black lines are reconfigurations. 
        }\vspace{0.1cm}
            \label{stepping-utility-a}
        \end{subfigure}
    \begin{subfigure}{0.5\textwidth}
                \includegraphics[width=1.0\columnwidth]{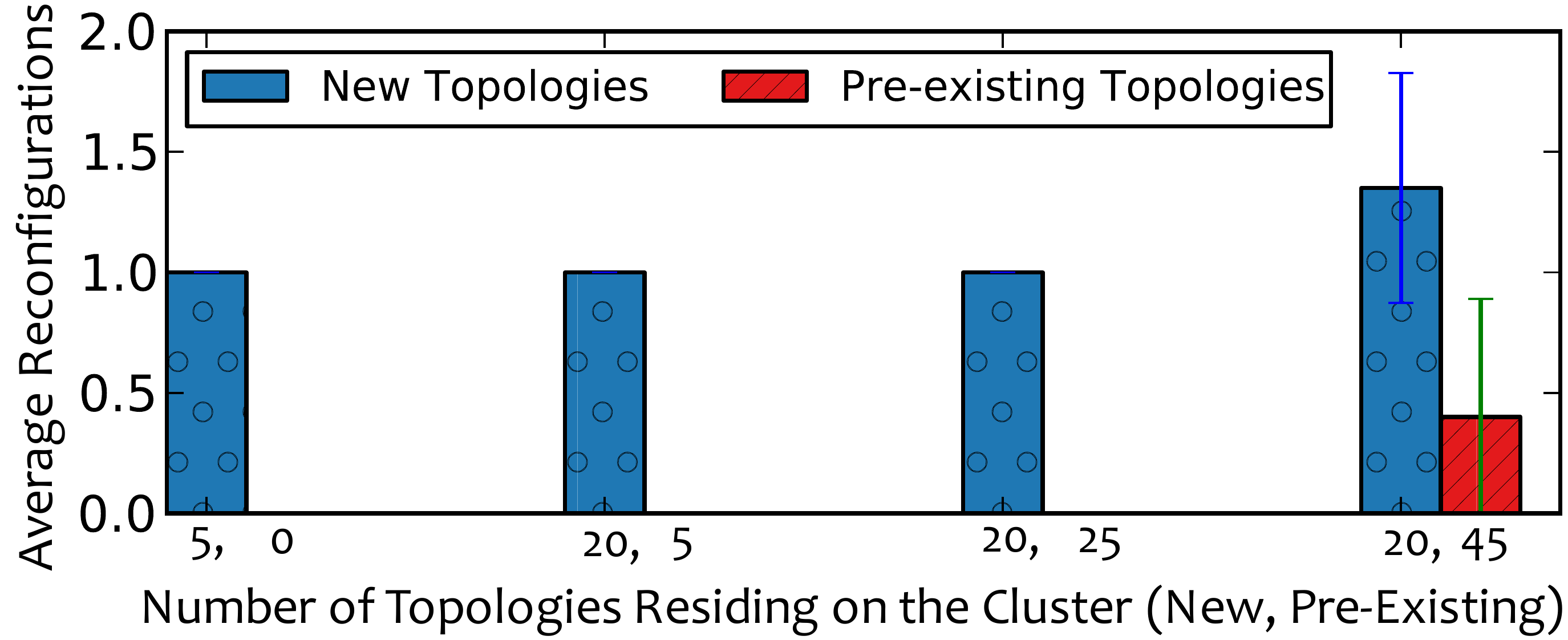}
                \vspace{-0.2cm}
        \caption{\it Average number of reconfigurations that must take place when new topologies are added to the cluster.} 
            \label{stepping-utility-b}
        \end{subfigure}
        \vspace{-0.3cm}
		\caption{{\bf Scalability w.r.t. No. of Topologies: } \it  Cluster has 5 tenants. 20 tenants are added every 2 hours until the 8 hour mark. 
		 }
		\label{stepping-utility}
	\end{center}
\vspace{-0.3cm}
\end{figure}

\noindent{\bf Increasing the Number of Topologies:} Fig.~\ref{stepping-utility} stresses Henge by overloading the cluster with  topologies over time. We start with a cluster of 5 PageLoad topologies, each with a latency SLO of 70 ms, and max utility value of 35.  Every 2 hours, we add 20 more  PageLoad topologies. 

Henge stabilizes better when there are more topologies. In the first 2 hours, Fig.~\ref{stepping-utility-a} shows that the average utility of the topologies is below the max, because Henge has less state space to maneuver with fewer topologies. 20 new tenant topologies at the 2 hour mark cause a large drop in  average utility but also open up the state space more--Henge quickly improves system utility to the max value. At the 4 hour and 6 hour marks, more topologies arrive. Henge stabilizes to max utility in both cases. 

Topologies arriving at the 8 hour mark cause contention. In Fig.~\ref{stepping-utility-a}, the average system utility drops not only due to the performance of the new tenants, but also because the pre-existing tenants are hurt. Henge converges both types of topologies, requiring fewer reconfiguration rounds for the pre-existing topologies (Fig.~\ref{stepping-utility-b}).

\begin{figure}[h!]
	\includegraphics[width=1.0\columnwidth]{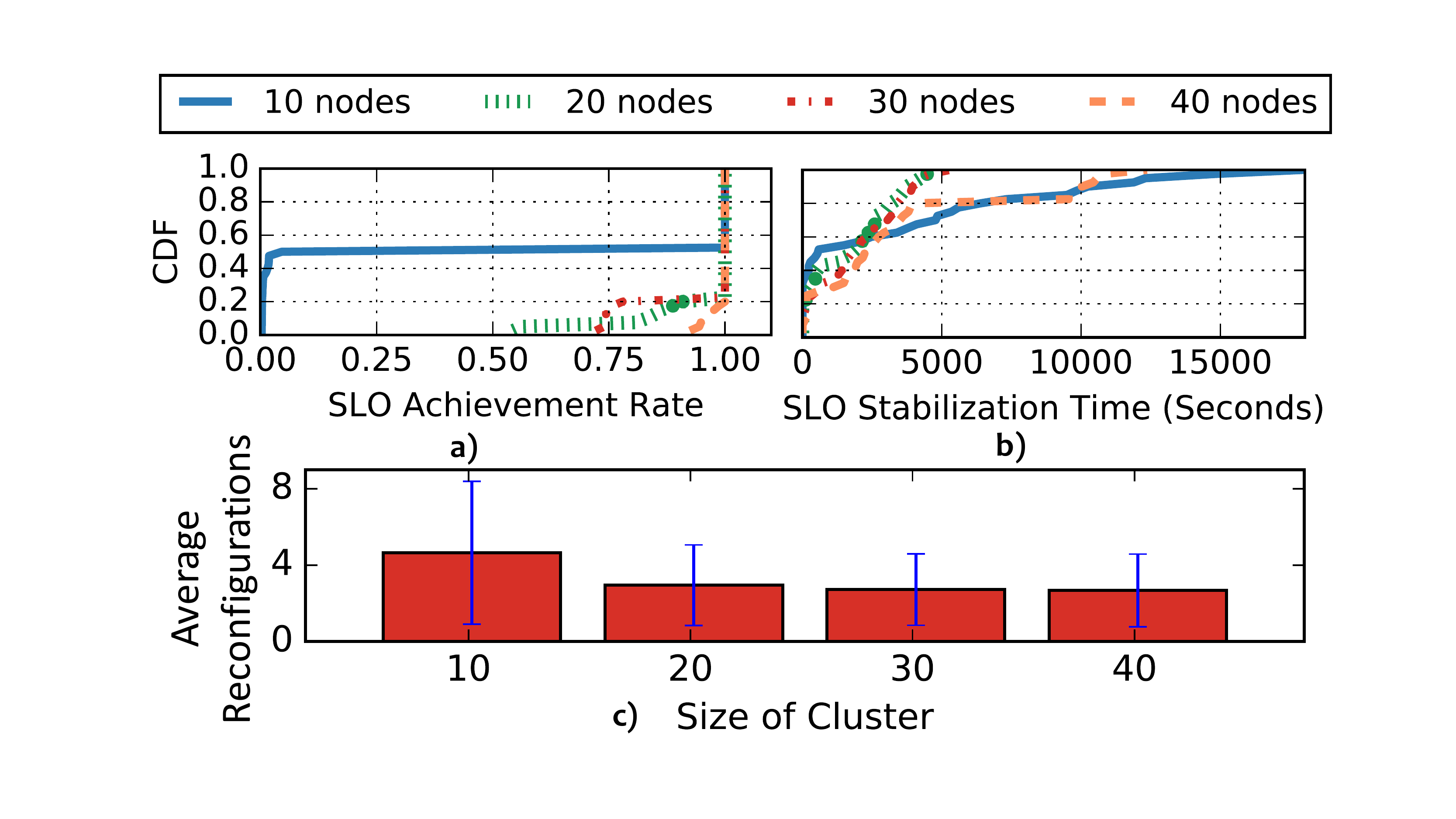}
	\vspace{-0.7cm}
		\caption{
		{\bf Scalability w.r.t No. of Machines: }
		\emph{ 40 jobs run on cluster sizes increasing from 10  to 40 nodes. a) CDF of  jobs according to fraction of SLO thresholds reached. b) CDF of convergence time. c) No. of reconfigurations until convergence.}
		}
		\label{scalability}

\vspace{-0.3cm}
\end{figure}

\noindent{\bf Increasing Cluster Size:} 
In Fig.~\ref{scalability} we run 40 topologies on clusters  ranging from 10 to 40 nodes. The machines  have two 2.4 GHz 64-bit 8-core processors, 64 GB RAM, and a 10 Gbps network. 20 topologies are PageLoad with latency SLOs of 80 ms and max utility 35. Among the rest, 8 are diamond topologies, 6 are star topologies and  6 are linear topologies, with juice SLOs of 1.0 and max utility 5.

From Fig.~\ref{scalability}a, Henge is able to provide SLO satisfaction for 40\% of the tenants even in an overloaded cluster of 10 nodes. As expected, in large clusters Henge has more room to maneuver and meets more SLOs. This is because CPUs saturate slower in larger clusters. In an overloaded cluster of 10 nodes, topologies at the 5th percentile are able to achieve only 0.3\% of their max utility. On the other hand, in clusters with 20, 30, and 40 machines, 5th percentile SLO satisfactions are higher: 56.4\%, 74.0\% and 94.5\% respectively. 



Fig.~\ref{scalability}b shows the time taken for topologies to converge to their highest utility.  Interestingly, while the 10 node cluster has a longer tail than 20 or 30 nodes, it converges faster at the median (537.2 seconds). 
Topologies at the tail of both the 10 and 40 node clusters take a longer time to converge. This is because in the 10 node cluster, greater reconfiguration is required per topology as there is more resource contention (see Fig.~\ref{scalability}c). At 40 nodes, collecting cluster information from  Nimbus daemons leads to a bottleneck. This can be alleviated by decentralized data gathering (beyond our scope). 


Fig.~\ref{scalability}c shows that the number of reconfigurations needed to converge is at most 2 $\times$ higher when resources are limited and does not otherwise vary with cluster size. Overall, Henge's performance generally improves with cluster size, and overheads are scale-independent. 

\subsubsection{Fault Tolerance}

\begin{figure}[h!]
 	\vspace{-0.1cm}	
	\begin{center}
\includegraphics[width=1.0\columnwidth]{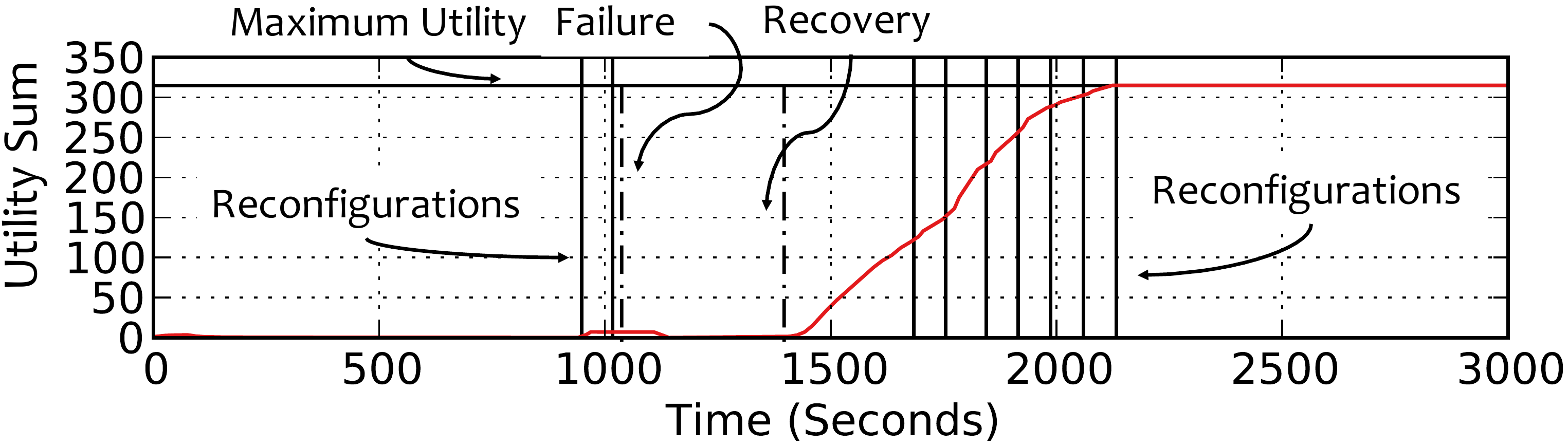}\vspace{-0.3cm}
\caption{{\bf Fault-tolerance:} \it Failure occurs at t=1020s, and recovers at t=1380 s. Henge makes no wrong decisions due to the failure, and immediately  converges to the max system utility after recovery. }
		\vspace{-0.5cm}
		\label{fault-tolerance}
	\end{center}

\end{figure}

Henge reacts gracefully to failures. In Fig.~\ref{fault-tolerance}, we run 9 topologies each with 70 ms SLO and 35 max utility. We introduce a failure at the worst possible time: in the midst of Henge's reconfiguration operations, at 1020 s. This severs communication between Henge and all the worker nodes; and Henge's Statistics module  is unable to obtain fresh information about jobs. We observe that Henge reacts conservatively by avoiding reconfiguration in the absence of data.  
At 1380 s, when communication is restored, Henge collects performance data for the next 5 minutes (until 1680 s) and then proceeds with reconfigurations as usual, meeting all SLOs. 

\subsection{Memory Utilization}
\label{sec:memory-utilization}

\begin{figure}[ht]
	\begin{center}
\includegraphics[width=1.0\columnwidth]{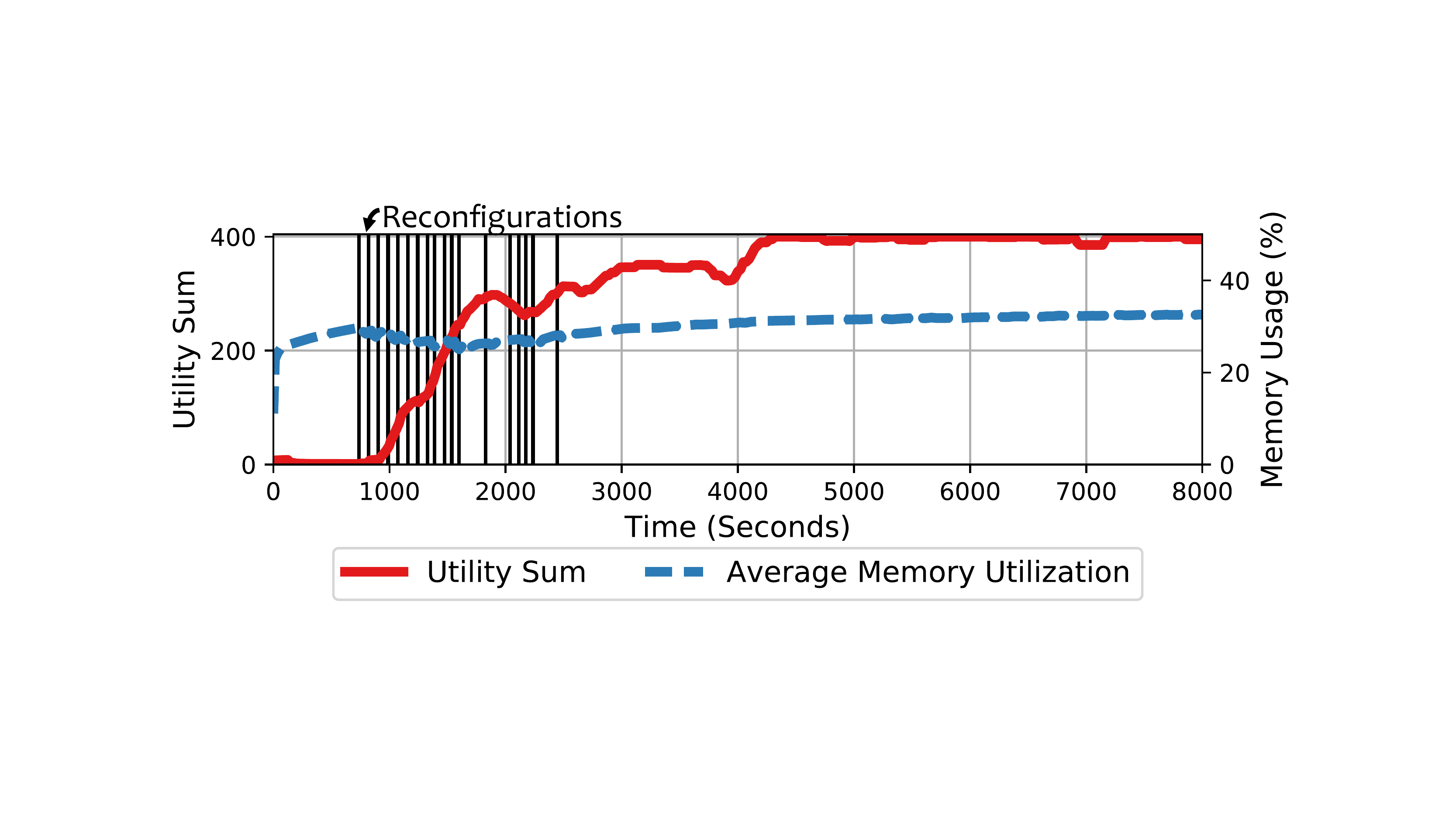}
\vspace{-0.75cm}
\caption{{\bf Memory Utilization:} \emph{ 8 jobs with joins and 30 s tuple retention.}}   
		\vspace{-0.5cm}
		\label{memory-utilization}
	\end{center}
\end{figure}

Fig.~\ref{memory-utilization} shows a Henge cluster with 8 memory-intensive topologies. Each topology has a max utility value of 50 and a latency SLO of a 100 ms. These topologies have join operations, and tuples are retained for 30 s, creating memory pressure at some cluster nodes. As the figure shows, Henge reconfigures memory-bound topologies quickly to reach total max utility of 400 by 2444s, and keeps  average memory usage below 36\%. Critically, the memory utilization (blue dotted line) plateaus in the converged state, showing that Henge is able to handle memory-bound topologies gracefully.

\vspace{-0.3cm}
\section{Related Work}
\label{related-work}


\noindent {\bf Elastic Stream Processing Systems:}
Traditional query-based stream processing systems such as Aurora~\cite{aurora} and Borealis~\cite{borealis} provide load-balancing~\cite{load-in-borealis} but not first-class multi-tenant elasticity. 
Modern general-purpose stream processing systems~\cite{flink,flume,samza,storm_web,s4,akidau2013millwheel,heron,naiad} do not natively handle adaptive elasticity. Ongoing work~\cite{spark-elasticity}  on Spark Streaming~\cite{spark_streaming} allows scaling but does not apply to resource-limited multi-tenant clusters.~\cite{seep,timestream} scale out stateful operators and  checkpoint, but do not scale in or support multi-tenancy. 


Resource-aware elasticity in stream processing~\cite{balkesen2013adaptive,cervino2012adaptive,fu2015drs,kalyvianaki2012overload,kleiminger2011balancing,satzger2011esc,peng2015r}   assumes infinite resources that the tenant can scale out to. 
~\cite{aniello2013adaptive,elastic-scaling-system-S,jain2006design,li2015supporting,stormy,schneider2009elastic,wu2007challenges} propose  resource provisioning but not multi-tenancy. Some works have focused on balancing load~\cite{twochoices,morethantwochoices}, optimal operator placement~\cite{SQPR,network-aware-placement,heinze2014latency} and scaling out strategies~\cite{heinze2014auto,heinze2015online} in stream processing systems. These approaches can be used in complement with Henge in various ways. \cite{heinze2014latency,heinze2014auto,heinze2015online} look at single-job elasticity, but not multi-tenancy.


Themis'~\cite{kalyvianaki2016themis} SIC 
metric is similar to juice, but Themis uses SIC to drop tuples. Henge does not drop tuples. 
Dhalion~\cite{floratou2017self} supports throughput SLOs for Heron, but it is unclear how this generalizes to varying input rates. It uses topology backpressure as a trigger for scaling out topologies, but because backpressure takes a while to propagate (e.g., after a spike), this approach is less responsive than using CPU load. 


\noindent {\bf Multi-tenant Resource Management Systems:} Resource schedulers like  YARN~\cite{yarn} and Mesos~\cite{mesos} can be run under stream processing systems, and manually tuned~\cite{cloudera:yarn-tune}.
Since the job internals are not exposed to the scheduler (jobs  run on containers) it is impossible to make fine-grained decisions for stream processing jobs in an automated fashion. 

\noindent {\bf Cluster Scheduling: } Some systems propose scheduling solutions to address resource fairness and SLO achievement~\cite{delimitrou2013paragon, quasar, ghodsi2011dominant, mace2015retro, rameshan2016hubbub, shue2012performance}. 
VM-based scaling approaches~\cite{vm-based} do not map directly and efficiently to expressive frameworks like stream processing systems. 
Among multi-tenant stream processing systems, Chronostream~\cite{wu2015chronostream} achieves elasticity through  migration across nodes. It does not support SLOs.  


\noindent {\bf  SLAs/SLOs in Other Areas:} 
SLAs/SLOs have been explored in other areas. 
Pileus~\cite{terry2013consistency} is a geo-distributed storage system that supports multi-level SLA requirements dealing with latency and consistency. 
Tuba~\cite{ardekani2014self} builds on Pileus and performs reconfiguration to adapt to changing workloads. 
SPANStore~\cite{wu2013spanstore} is a geo-replicated storage service that automates trading off cost vs. latency, while being consistent and fault-tolerant. 
E-store~\cite{e-store}  re-distributes hot and cold data chunks across nodes in a cluster if load exceeds a threshold. Cake~\cite{wang2012cake} supports latency and throughput SLOs in multi-tenant storage settings.

\vspace{-1.0em}
\section{Conclusion}
\label{conclusion}

We presented Henge, a system for intent-driven (SLO-based) multi-tenant stream processing. Henge provides SLO satisfaction for topologies (jobs) with  latency and/or throughput requirements. To make throughput SLOs independent of input rate and topology structure, Henge uses a new  relative throughput  metric called juice.  In a cluster, when jobs  miss their SLO, Henge uses three kinds of actions (reconfiguration, reversion or reduction) to improve the sum utility achieved by all jobs throughout the cluster. Our experiments with real Yahoo! topologies and Twitter datasets have shown that in multi-tenant settings with a mix of SLOs, Henge: 
i) converges quickly to max system utility when resources suffice; 
ii) converges quickly to a high system utility when the cluster is constrained; 
iii) gracefully handles dynamic workloads, both abrupt (spikes, natural fluctuations) and gradual (diurnal patterns, Twitter datasets);
iv) scales gracefully with cluster size and number of jobs; and
v) is failure tolerant. 

\bibliographystyle{acm}
\bibliography{bib/kalim-eurosys17}

\end{document}